\documentclass[
 reprint,
 superscriptaddress,
 amsmath,amssymb,
 aps,
 prb,
]{revtex4-2}

\usepackage{graphicx}
\usepackage{dcolumn}
\usepackage{bm}
\usepackage{mathtools}
\usepackage{float}
\usepackage[usenames,dvipsnames]{color}
\usepackage[table]{xcolor}
\usepackage[
    colorlinks=true,
    citecolor=blue,
    linkcolor=blue,
    urlcolor=blue
]{hyperref}
\usepackage{cleveref}
\usepackage[nolist]{acronym}
\usepackage{physics}
\usepackage{booktabs}
\usepackage{multirow}
\usepackage{array}
\usepackage{tabularray}
\usepackage{nicematrix}
\usepackage{bbold}
\usepackage{siunitx}

\usepackage{ulem}

\newcommand{\affiliationBremen}{
Institute for Theoretical Physics and Bremen Center for Computational Materials Science,
University of Bremen, 28359 Bremen, Germany%
}

\newcommand{\affiliationMPSD}{
Max Planck Institute for the Structure and Dynamics of Matter,
Center for Free-Electron Laser Science (CFEL),
Luruper Chaussee 149, 22761 Hamburg, Germany%
}

\newcommand{\affiliationCCQ}{
Center for Computational Quantum Physics (CCQ), Flatiron Institute,
162 Fifth Avenue, New York, NY 10010, USA%
}

\begin{document}


\title{Microscopic mechanism for resonant light-enhanced pair correlations in K$_3$C$_{60}$}

\author{Juan~I. Aranzadi}
\email{aranzadi@uni-bremen.de}
\affiliation{\affiliationBremen}

\author{Joseph Tindall}
\affiliation{\affiliationCCQ}

\author{Paul Fadler}
\affiliation{\affiliationBremen}

\author{Michael~A. Sentef}
\email{sentef@uni-bremen.de}
\affiliation{\affiliationBremen}
\affiliation{\affiliationMPSD}

\date{\today}

\begin{abstract}
Recent experiments on K$_3$C$_{60}$ revealed a giant enhancement of the light-induced superconducting-like optical response for pump frequencies near \SI{10}{\tera\hertz}, with an efficiency roughly two orders of magnitude larger than for off-resonant excitation. Here we show that a resonant enhancement of pair correlations arises naturally in a driven purely electronic model of K$_3$C$_{60}$ with \textit{ab initio} parameters. The underlying mechanism is a symmetry-constrained two-photon pathway: the first photon drives the system from the even-parity ground state to an intermediate odd-parity manifold, while the second photon induces a transition to an even-parity excited state with enhanced pair correlations. Larger-cluster calculations show that the corresponding resonance energy is strongly renormalized downward with system size and connectivity, reflecting the kinetic-energy gain of delocalized photo-excited doublon-holon configurations. A simplified single-orbital model reproduces this scaling trend and reaches a 14-site fcc cluster, where the resonant peak is pushed to $\sim\SI{30}{\tera\hertz}$, with a trend compatible with a further reduction toward the experimental \SI{10}{\tera\hertz} scale in larger systems. Varying the Hubbard coupling strength, we find that the resonance is lowest and the pairing enhancement strongest near intermediate couplings, where doublon-holon excitations are both well defined and mobile. Our results establish a purely electronic mechanism for resonant light-enhanced pair correlations in K$_3$C$_{60}$ and support the interpretation of the experimental \SI{10}{\tera\hertz} resonance as optical access to a paired many-body excited state rather than improved metallicity. More broadly, they suggest that related resonant pathways may arise in other intermediate-coupling Hubbard materials.
\end{abstract}

\maketitle

\section{Introduction}

A major goal of ultrafast quantum materials science is to induce nontrivial, far-from-equilibrium quantum many-body states of matter~\cite{basov_towards_2017,de_la_torre_colloquium_2021}, including states with no known equilibrium counterpart. Among the most prominent examples are superconducting-like optical responses observed far above the equilibrium transition temperature $T_c$ following mid-infrared excitation~\cite{cavalleri_photo-induced_2018}. First reported in a stripe-ordered cuprate~\cite{fausti_light-induced_2011}, such signatures were subsequently observed in underdoped YBa$_2$Cu$_3$O$_{6+x}$~\cite{hu_optically_2014,kaiser_optically_2014,mankowsky_nonlinear_2014,hunt_dynamical_2016}, organic $\kappa$-(BEDT-TTF)$_2X$ salts~\cite{buzzi_photomolecular_2020,buzzi_phase_2021}, and K$_3$C$_{60}$~\cite{mitrano_possible_2016}. More recently, the observation of transient magnetic-field expulsion in optically driven YBa$_2$Cu$_3$O$_{6.48}$ has provided further evidence for the superconducting-like character of the photo-induced state~\cite{fava_magnetic_2024}. Notably, these materials are all correlated electron systems in which interaction and kinetic-energy scales compete. Their equilibrium normal states above $T_c$ also display unconventional behavior, including pseudogap physics and stripe order in the cuprates and pronounced superconducting fluctuations or precursor correlations in the $\kappa$-salts and K$_3$C$_{60}$~\cite{buzzi_phase_2021,Jotzu2023}. Uemura has proposed a unified interpretation of these equilibrium and photoinduced responses in terms of preformed pairs, local phase coherence, and competition between superconducting and magnetic order~\cite{Uemura2019}. In K$_3$C$_{60}$, the light-induced state has furthermore been linked to long-lived metastable behavior and quasi-dc transport measurements~\cite{budden_evidence_2021}, with pressure dependence supporting a superconducting interpretation over simple enhanced metallicity~\cite{cantaluppi_pressure_2018}. Most strikingly, a recent pump-frequency study uncovered a sharp resonance near \SI{10}{\tera\hertz} ($\sim\SI{41}{\milli\electronvolt}$), where the photo-susceptibility increases by about two orders of magnitude relative to off-resonant excitation~\cite{rowe_resonant_2023}.

The sharp \SI{10}{\tera\hertz} resonance provides a key clue to the microscopic origin of light-enhanced pairing in K$_3$C$_{60}$. While many theoretical scenarios have been proposed for light-induced superconductivity in fullerides~\cite{Kim_2016,Knap_2016,Murakami_2017,Babadi_2017,Mazza_2017,Kennes2017,Dasari_2018,nava_cooling_2018,Buzzi_2021,Chattopadhyay2025,sous2025,Grankin_2026,chattopadhyay2026}, the origin of this very sharp resonant enhancement has remained unresolved to date. Here we show that an \textit{ab initio}-motivated Hubbard--Kanamori model supports a symmetry-constrained two-photon transition into a high-energy even-parity state with strongly enhanced pair correlations. As the system size increases, the excitation energy of this state is renormalized downward because the pumped doublon delocalizes across the cluster and gains kinetic energy. This identifies a purely electronic route to resonant light-enhanced pairing correlations in K$_3$C$_{60}$, not directly involving phonons. Moreover, we find that the paired excited state exists even in a single-orbital Hubbard model. Therefore, the mechanism points to a wider class of intermediate-coupling Hubbard materials with $U$ and bandwidth $W$ on comparable scales---including cuprates, nickelates, or organic superconductors such as $\kappa$-(BEDT-TTF)$_2X$---as natural systems in which related resonant pathways may occur in suitable regions of their phase diagrams.

\section{Model}

The equilibrium low-energy physics of K$_3$C$_{60}$ is well captured by three half-filled $t_{1u}$ bands with local Hubbard--Kanamori interactions and an inverted Hund coupling arising from Jahn--Teller screening~\cite{capone2009,nomura2012_abinitio,nomura_unified_2015}. The low-energy Hamiltonian can be written as $H_{\text{K$_3$C$_{60}$}}=H_{\mathrm{local}}+H_{\mathrm{hop}}$, where the local interactions are
\begin{equation}
    \begin{split}
        H_{\mathrm{local}}={}&
        U\sum_{\alpha}
        \hat{n}_{\alpha,\uparrow}\hat{n}_{\alpha,\downarrow}
        \\
        &+
        \left(U-2J_{\mathrm{inv}}\right)
        \sum_{\alpha\neq\beta}
        \hat{n}_{\alpha,\uparrow}\hat{n}_{\beta,\downarrow}
        \\
        &+
        \left(U-3J_{\mathrm{inv}}\right)
        \sum_{\alpha<\beta,\sigma}
        \hat{n}_{\alpha,\sigma}\hat{n}_{\beta,\sigma}
        \\
        &-
        J_{\mathrm{inv}}
        \sum_{\alpha\neq\beta}
        c^\dagger_{\alpha,\uparrow}c_{\alpha,\downarrow}
        c^\dagger_{\beta,\downarrow}c_{\beta,\uparrow}
        \\
        &+
        J_{\mathrm{inv}}
        \sum_{\alpha\neq\beta}
        c^\dagger_{\alpha,\uparrow}c^\dagger_{\alpha,\downarrow}
        c_{\beta,\downarrow}c_{\beta,\uparrow},
    \end{split}
\end{equation}
and the hopping term reads
\begin{equation}
    \begin{split}
        H_{\mathrm{hop}}={}&
        \sum_{i\neq j}\sum_{\alpha,\beta,\sigma}
        T_{\alpha,\beta}\!\left(\mathbf{R}_i-\mathbf{R}_j\right)
        \\
        &\times
        \left(
        c^\dagger_{i,\alpha,\sigma}c_{j,\beta,\sigma}
        +\mathrm{h.c.}
        \right).
    \end{split}
    \label{eq:hubb-kana}
\end{equation}
Here, $T_{\alpha,\beta}(\mathbf{R}_i-\mathbf{R}_j)$ denotes the transfer matrix for hopping between sites separated by $\mathbf{R}_i-\mathbf{R}_j$. We use \textit{ab initio}-derived transfer parameters from Ref.~\cite{nomura2012_abinitio} (see Supplemental Material, Sec.~\ref{sec:sup_hamiltonian}, for explicit values), together with $U=\SI{500}{\milli\electronvolt}$ and $J_{\mathrm{inv}}=\SI{-20}{\milli\electronvolt}$~\cite{nomura2012_abinitio,nomura_2015_abinitio}, which gives $U/W\simeq1$.

The system is driven by a spatially homogeneous electric field along the [100] direction of the fcc lattice, coupling through the cluster-dipole operator,
\begin{equation}
    \begin{split}
        H_{\mathrm{drive}}(t)
        &=
        \mathbf{E}(t)\cdot\mathbf{D}
        \\
        &=
        E\sin(2\pi\nu t)D^x
        \\
        &=
        E\sin(2\pi\nu t)
        \sum_{i,\alpha}R_i^{(x)}n_{i,\alpha}.
    \end{split}
    \label{eq:drive}
\end{equation}
Here, $E$ is the field amplitude, $\nu$ is the drive frequency, and we choose the drive direction to be $x$ without loss of generality. To quantify light-enhanced pairing, we monitor the observable
\begin{equation}
    \hat{P}
    =
    \sum_{i\neq j}\sum_{\alpha,\beta}
    c^\dagger_{i\alpha\uparrow}
    c^\dagger_{i\alpha\downarrow}
    c^{\phantom{\dagger}}_{j\beta\downarrow}
    c^{\phantom{\dagger}}_{j\beta\uparrow},
\end{equation}
which measures nonlocal pair correlations between local doublons~\footnote{In an $s$-wave, spin-singlet superconductor, the pair-field susceptibility corresponding to $\hat{P}$ would diverge at the critical temperature.}

\begin{figure}[t]
    \centering
    \includegraphics[width=\columnwidth]{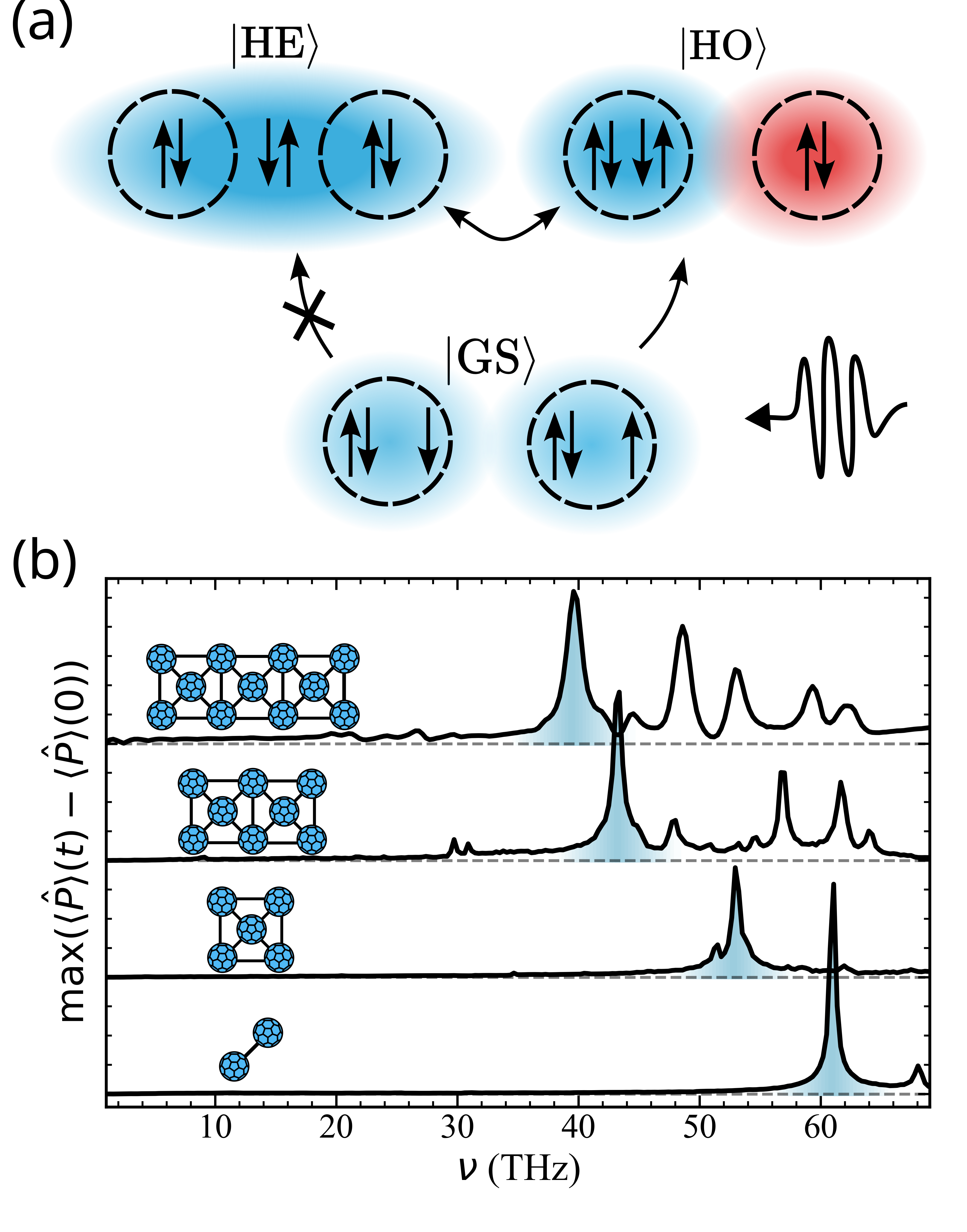}
    \caption{
    \textbf{Two-photon pathway and resonant enhancement of pairing correlations.}
    \textbf{(a)} Schematic excitation pathway. The even-parity inverted-Hund's-rule ground state (GS) couples by one-photon absorption to an odd-parity high-energy state (HO), which in turn couples to a high-energy even-parity state (HE) with enhanced pairing. Direct GS$\to$HE coupling is symmetry forbidden.
    \textbf{(b)} Light-induced enhancement of pair correlations $\max_t[\langle\hat P\rangle(t)-\langle\hat P\rangle(0)]$ as a function of laser frequency $\nu$, obtained from full time evolution under the drive described in Eq.~\ref{eq:drive}, for the 2-, 5-, 8-, and 11-site 2D clusters depicted alongside each curve. Curves are offset vertically and ordered by increasing cluster size from bottom to top. The peak highlighted in blue corresponds to the two-photon transition described in the main text.
    }
    \label{fig:main}
\end{figure}

\section{Methods}

We begin with exact diagonalization on small clusters, where the full spectrum and dipole matrix elements can be resolved explicitly. Already for a buckyball dimer, the induced pair correlation exhibits a pronounced sub-$U$ resonance around \SI{61}{\tera\hertz} (\SI{252}{\milli\electronvolt}), well below the bare interaction scale. Analysis of the spectrum shows that the dynamics are governed by a small set of states selected by inversion symmetry and dipole matrix elements; see Supplemental Material, Sec.~\ref{sec:sup_drivesym}. Because the undriven Hamiltonian is inversion symmetric, its eigenstates have definite parity. The dipole operator is odd under inversion and therefore couples only states of opposite parity. Since the ground state is even, the relevant high-pairing even state cannot be reached directly by one-photon absorption. Instead, the drive first populates an odd-parity excited manifold virtually and from there accesses a high-energy even-parity state with one additional delocalized doublon and strongly enhanced pair correlations, as illustrated in Fig.~\ref{fig:main}(a).

This motivates a minimal three-state picture consisting of an even ground state $|\mathrm{GS}\rangle$, an odd intermediate state $|\mathrm{HO}\rangle$, and a high-pairing even state $|\mathrm{HE}\rangle$. For the buckyball dimer, both excitation energies are of order $U$, with $E_{\mathrm{HE}}-E_{\mathrm{GS}}\simeq E_{\mathrm{HO}}-E_{\mathrm{GS}}$. The resonant condition is therefore
\begin{equation}
    2h\nu\simeq E_{\mathrm{HE}}-E_{\mathrm{GS}},
\end{equation}
corresponding to a two-photon transition into the even-parity target state mediated by the odd sector. The key point is that the resonance is set not by a conventional bright excitation, but by symmetry-constrained optical access to an even-parity state with strongly enhanced pair correlations. This three-state truncation already reproduces the exact-diagonalization dynamics of the pairing observable with high accuracy; see Supplemental Material, Secs.~\ref{sec:two-bucky-ED} and~\ref{sec:FGR}. This shows that the resonance is controlled by a remarkably small, physically selected subset of the Hilbert space.

The full Hilbert space grows prohibitively fast with system size, making brute-force dynamics impossible beyond very small clusters. We therefore exploit the coupling structure identified above to construct an effective reduced basis for the driven problem. We first obtain the many-body ground state using DMRG in matrix-product-state form and then build a Krylov subspace $\mathcal{K}_n(H_0,D^x|\mathrm{GS}\rangle)$ that targets the odd sector directly coupled to it. From this space, we retain the dominant odd states and, for each such state, generate an even-sector Krylov space after a second application of the dipole operator. This DMRG+Krylov construction yields an effective Hamiltonian tailored to the two-photon dynamics while retaining the underlying microscopic model. We implement this algorithm using the ITensor library~\cite{itensor}; see Supplemental Material, Sec.~\ref{sec:DMRG+Krylov}, for details.

\section{Results}

\subsection{Finite-size scaling and effective one-orbital description}

We first apply this construction to 2D stripe clusters. As shown in Fig.~\ref{fig:main}(b), the main resonance in the induced pair correlation shifts systematically downward with size: from \SI{61}{\tera\hertz} for the dimer to \SI{53}{\tera\hertz} for five sites, \SI{43}{\tera\hertz} for eight sites, and \SI{40}{\tera\hertz} for eleven sites. This trend shows that the relevant excited state is not a purely local doublon excitation at energy $U$, but an extended many-body excitation whose energy is lowered as the extra doublon delocalizes across the cluster and gains kinetic energy. At the same time, this coherent delocalization enhances pair correlations relative to the ground state.

\begin{figure}[t]
    \centering
    \includegraphics[width=\columnwidth]{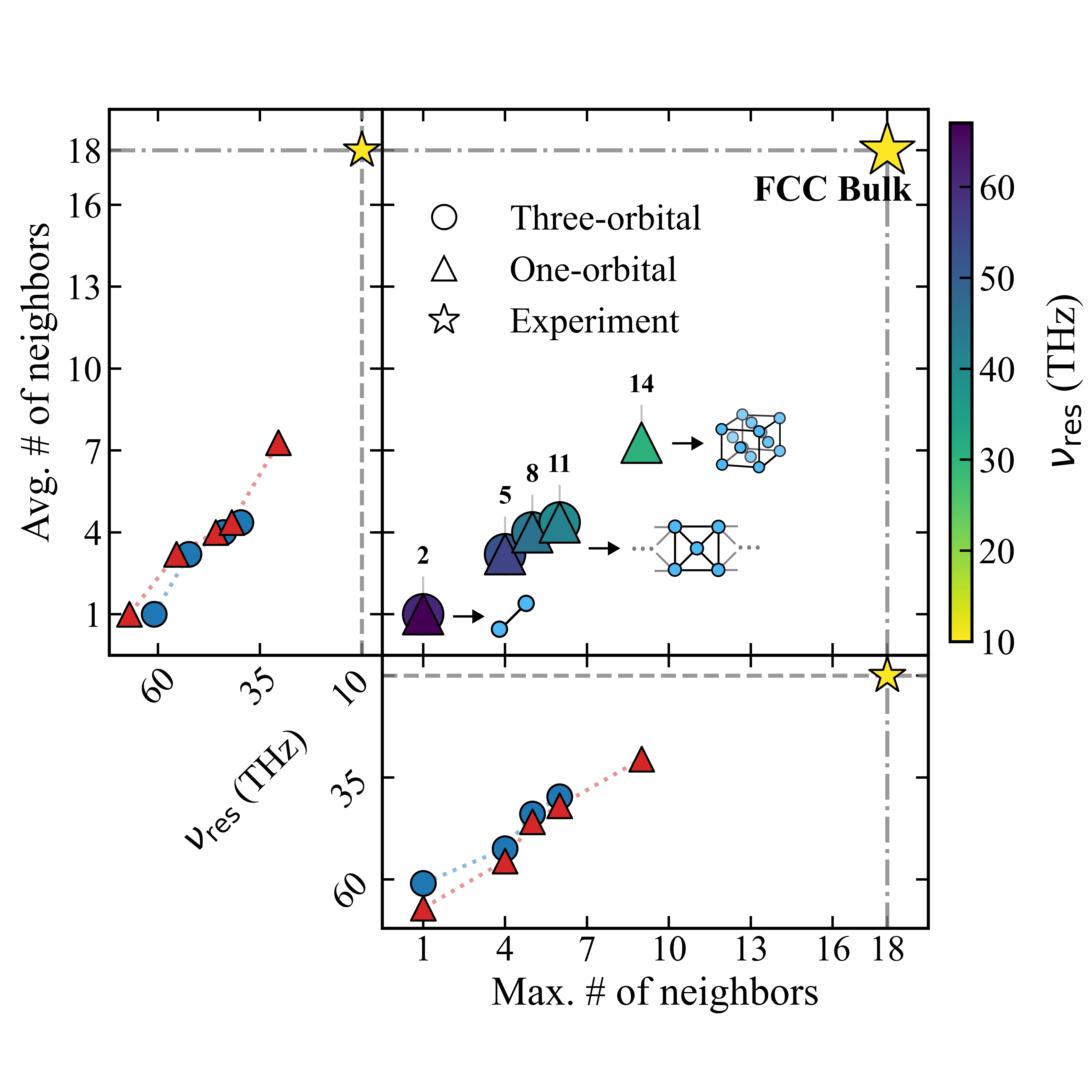}
    \caption{
    \textbf{Scaling of the two-photon resonance with cluster connectivity.}
    Central panel: average versus maximum number of neighbors for the cluster geometries considered, where the number of neighbors of a site counts all nearest and next-nearest neighbors. Circles (triangles) denote the three-orbital (one-orbital) model, the marker color encodes the resonance frequency $\nu_{\mathrm{res}}$, and the number above each marker gives the number of sites in the cluster; representative geometries are sketched alongside. Left and bottom panels: $\nu_{\mathrm{res}}$ as a function of the average and maximum number of neighbors, respectively. Both models show a consistent downward renormalization of $\nu_{\mathrm{res}}$ with increasing connectivity, reflecting the kinetic-energy gain from doublon-holon delocalization. The dotted gray line marks the experimentally observed \SI{10}{\tera\hertz} resonance~\cite{rowe_resonant_2023}, and the dash-dotted gray line marks the fcc bulk coordination number of 18, including nearest and next-nearest neighbors. Their intersection, corresponding to the experiment, is indicated by the star.
    }
    \label{fig:cluster_comparison}
\end{figure}

Despite the numerical simplifications enabled by the Krylov-subspace constructions, larger clusters still remain out of reach in the full three-orbital model. We therefore turn to an effective one-orbital model. The choice of this one-orbital Hubbard model is motivated by the fact that the inverted Hund coupling strongly favors low-spin configurations and allows us to isolate one orbital on each buckyball that is most relevant for the composition of the HE state, as discussed in more detail below. The corresponding effective Hamiltonian reads
\begin{equation}
    \begin{split}
        H_{\mathrm{1orb}}={}&
        U\sum_i\hat{n}_{i\uparrow}\hat{n}_{i\downarrow}
        \\
        &+
        \sum_{\langle i,j\rangle,\sigma}
        t_{\mathrm{NN}}
        \left(
        c^\dagger_{i\sigma}c_{j\sigma}
        +\mathrm{h.c.}
        \right)
        \\
        &+
        \sum_{\langle\!\langle i,j\rangle\!\rangle,\sigma}
        t_{\mathrm{NNN}}
        \left(
        c^\dagger_{i\sigma}c_{j\sigma}
        +\mathrm{h.c.}
        \right).
    \end{split}
\end{equation}
We choose $U=\SI{500}{\milli\electronvolt}$, $t_{\mathrm{NN}}=\SI{60}{\milli\electronvolt}$, and $t_{\mathrm{NNN}}=\SI{-9.4}{\milli\electronvolt}$, preserving the ratio $U/W\simeq1$ of the three-orbital model. We find that this one-orbital model reproduces the same resonant enhancement of pair correlations as the full model, and the resulting resonance frequencies follow a remarkably similar finite-size trend, as summarized in Fig.~\ref{fig:cluster_comparison}.

\begin{figure*}[t]
    \centering
    \includegraphics[width=\textwidth]{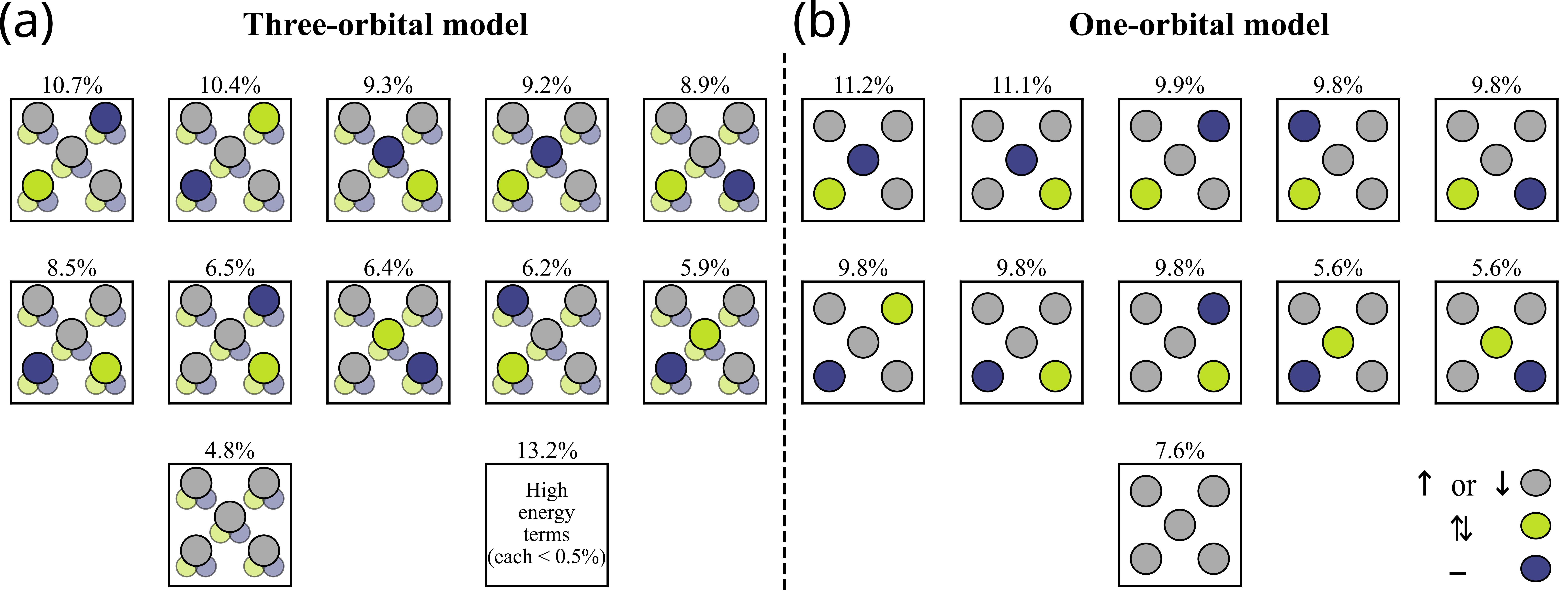}
    \caption{
    \textbf{Many-body state tomography of a representative high-energy even-parity state.}
    Probability distribution of the HE state in the doublon-holon-fermion configuration basis, with the spin degrees of freedom summed over, for the 5-site \textbf{(a)} three-orbital and \textbf{(b)} one-orbital clusters. The HE state is well captured by a superposition of the $N(N-1)$ possible arrangements of a doublon-holon pair across $N$ sites. Configurations related by inversion symmetry along $y$ are combined by summing their probabilities. The cluster $D_4$ symmetry is slightly broken due to the alignment of the drive along the $x$ direction.
    }
    \label{fig:tomography}
\end{figure*}

The reduced numerical complexity of the one-orbital model then allows us to go beyond the largest system size achievable for the three-orbital model and access a 3D 14-site cluster, which corresponds to the full fcc crystallographic unit cell. In this geometry, we find a pronounced resonant peak near \SI{30}{\tera\hertz}, demonstrating that the downward renormalization continues as both dimensionality and coordination number increase. This value remains above the experimental \SI{10}{\tera\hertz} scale, but Fig.~\ref{fig:cluster_comparison} shows that even a 14-site 3D cluster is still far from the thermodynamic bulk limit. In particular, it cannot capture the full kinetic-energy gain available to a fully delocalized doublon in a macroscopic crystal. The data therefore support a consistent physical picture: light-enhanced pairing arises from symmetry-constrained optical access to a highly paired excited state whose energy is lowered by delocalization.

We now turn to a more detailed understanding of why the one-orbital model is able to capture the relevant physics of the HE state. To this end, we show in Fig.~\ref{fig:tomography} the state tomography of the HE state responsible for the light-induced pair-correlation resonance for both the three- and one-orbital 5-site clusters. The relevant three-orbital HE state is dominated by a frozen doublon-holon background together with one additional doublon-holon pair that remains delocalized across the cluster. This extra-pair structure closely matches the corresponding one-orbital HE state. This similarity explains why the two models exhibit a nearly identical finite-size scaling of the resonance: in both cases, the relevant excitation is governed by the kinetic-energy gain associated with the delocalization of the additional doublon-holon pair.

\begin{figure}[t]
    \centering
    \includegraphics[width=\columnwidth]{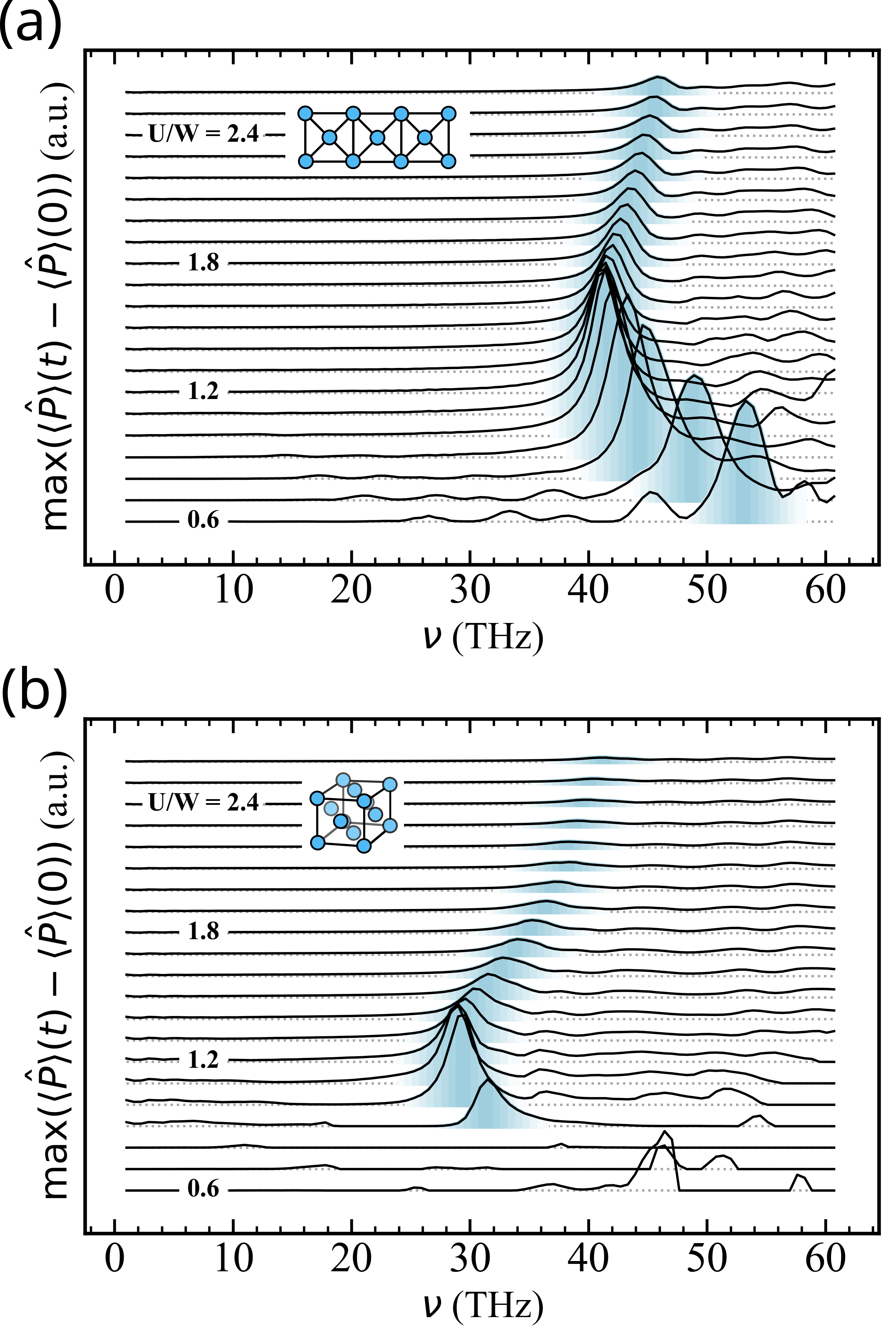}
    \caption{
    \textbf{Light-enhanced pair correlations as a function of $U/W$.}
    \textbf{(a)} 11-site 2D cluster and \textbf{(b)} 14-site 3D fcc cluster, shown in the insets. The vertical axis shows $\max_t[\langle\hat P\rangle(t)-\langle\hat P\rangle(0)]$, resolved as a function of laser frequency $\nu$, obtained from full time evolution under the drive described in Eq.~\ref{eq:drive}. Each trace corresponds to a different value of $U/W$, obtained by rescaling the hopping amplitudes at fixed $U=\SI{500}{\milli\electronvolt}$. Curves are offset vertically and ordered by increasing $U/W$ from bottom to top. The peak highlighted in blue corresponds to the two-photon transition described in the main text.
    }
    \label{fig:uw_sweep}
\end{figure}

\subsection{Dependence on correlation strength}

The observation that the single-orbital Hubbard model already captures the key physics of the light-induced paired state suggests that the effect is more general than the specific negative-$J$ model for K$_3$C$_{60}$. It is therefore natural to investigate systematically the role of the correlation strength in the HE-state resonance. We compute the light-enhanced pair correlations for varying interaction-to-bandwidth ratio $U/W$ in Fig.~\ref{fig:uw_sweep}, fixing $U=\SI{500}{\milli\electronvolt}$ and changing $W$ by uniformly rescaling the hopping parameters.

The $U/W$ spectral peak profile exhibits a nonmonotonic, nose-like shape, reaching the lowest resonant frequency near $U/W\approx1$. This feature arises from the competition between charge localization and kinetic-energy gain. In the Mott limit, $U/W\gg1$, the relevant HE states reside at the lower edge of the upper Hubbard band. Since the kinetic energy is strongly suppressed in this regime, the excitation energy is approximately $U$ and can be accessed via two photons of approximately $U/2$ energy. Conversely, in the weakly correlated metallic limit, $U/W\ll1$, the physics is dictated by the large kinetic-energy scale $W$, which proportionally pushes the highly paired excitations to higher energies. In this regime, the resonant peak vanishes entirely for the 14-site 3D cluster, likely because the doublon-holon excitation is no longer well defined and merges into the two-particle continuum.

It is precisely in the intermediate-coupling regime, $U/W\approx1$, that the system achieves an optimal balance. Here, the photo-excited doublon-holon pairs are well defined due to the interaction, yet they retain sufficient mobility to delocalize across the cluster. This delocalization provides a kinetic-energy gain that minimizes the required excitation energy of the HE state. The appearance of an optimal regime across different cluster geometries suggests that this interplay may constitute a more general mechanism for light-induced superconducting pairing in driven Hubbard models.

We note that the ground state of the 14-site 3D cluster undergoes a parity change from even to odd for $U/W\gtrsim1.16$, with the odd-even energy gap remaining below \SI{4}{\milli\electronvolt} in this $U/W$ range. As shown in Supplemental Material, Sec.~\ref{sec:sup_parity}, a time evolution initialized from the odd-parity ground state does not exhibit a resonant enhancement of pair correlations, consistent with the two-photon process requiring an even$\to$odd$\to$even transition sequence. For the simulations in Fig.~\ref{fig:uw_sweep}(b) at $U/W\gtrsim1.16$, we therefore initialize the dynamics from the lowest-energy even-parity state. This is physically justified because the experiments on K$_3$C$_{60}$ are performed at temperatures well above this gap, ensuring thermal access to both parity sectors. In all other clusters considered here, the ground state has even parity.

\subsection{Nonlinear terahertz spectroscopy}

\begin{figure*}[t]
    \centering
    \includegraphics[width=\textwidth]{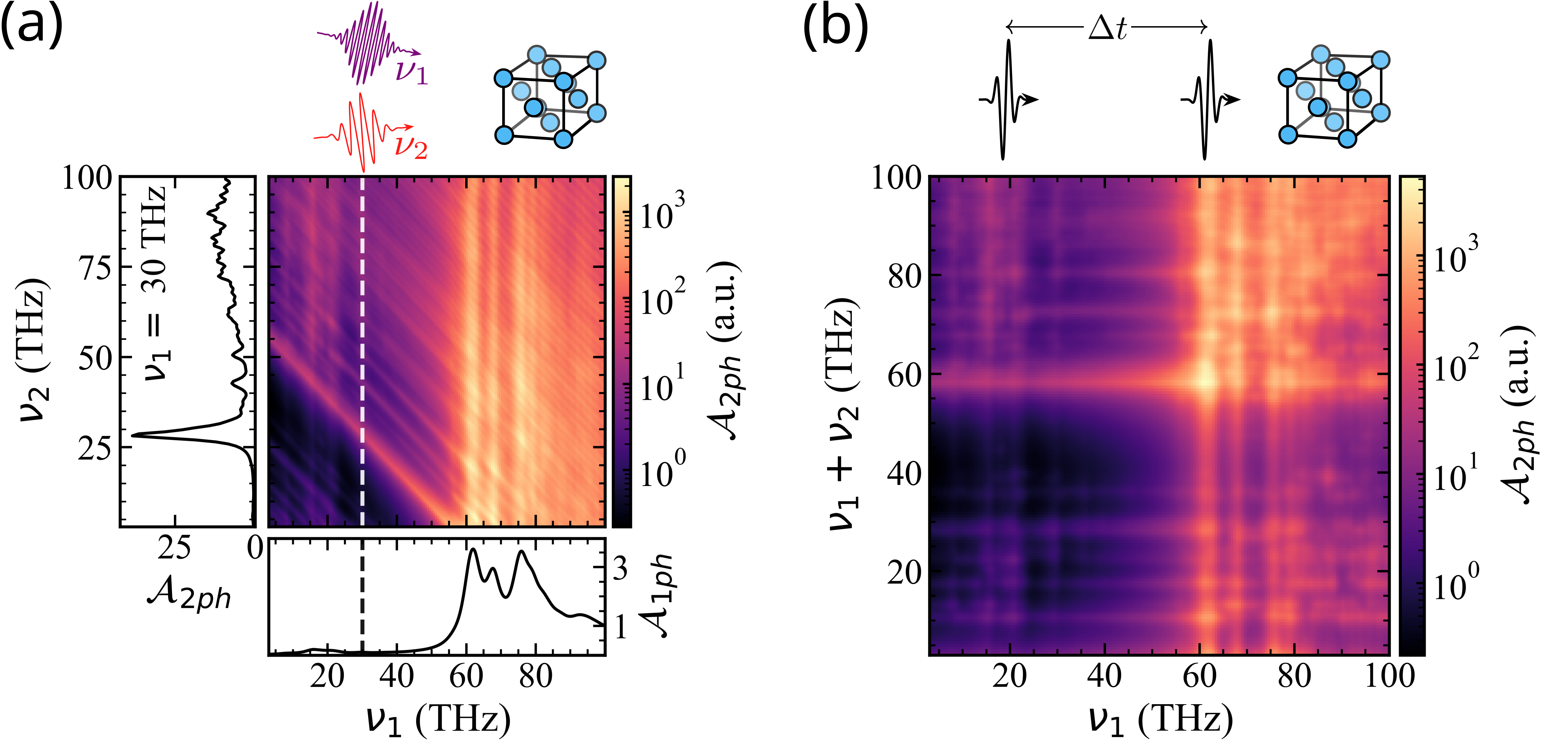}
    \caption{
    \textbf{Nonlinear terahertz spectroscopy.}
    \textbf{(a)} Two-photon absorption spectrum of the 14-site 3D fcc one-orbital cluster in the $(\nu_1,\nu_2)$ plane, a natural representation for a two-color drive. Bottom panel: one-photon absorption spectrum. Left panel: cut along the white dotted line. Vertical features are one-photon resonances, while antidiagonal features at fixed $\nu_1+\nu_2$ are two-photon transitions into even-parity states. The dominant antidiagonal at $\nu_1+\nu_2\approx\SI{60}{\tera\hertz}$ identifies the resonant pathway into the HE state. Under monochromatic driving, $\nu_1=\nu_2\equiv\nu$, this reduces to the $\nu\approx\SI{30}{\tera\hertz}$ pairing resonance.
    \textbf{(b)} Same spectrum reparametrized as $(\nu_1,\nu_1+\nu_2)$, a natural representation for a time-delay experiment, where the $y$ axis corresponds to the energy of the final even-parity state. Here, two-photon resonances are horizontal lines. Features below the $\nu_1=\nu_2$ diagonal reveal Raman-like pathways reaching lower even-parity states via stimulated emission from HO.
    }
    \label{fig:thz_spec}
\end{figure*}

We finally turn to how the proposed two-photon pathway for light-induced pairing could be probed spectroscopically. Nonlinear terahertz spectroscopy is a natural candidate method and has been used to probe collective modes in driven superconductors~\cite{shimano2014,katsumi2024,liu2025}. We compute the corresponding two-photon spectrum for the 14-site 3D fcc cluster, resolved as a function of two independent drive frequencies $\nu_1$ and $\nu_2$. The two-photon susceptibility follows from the imaginary part of the correlator
\begin{equation}
    G_{\mathrm{2ph}}
    =
    \left\langle
    \hat{D}
    \frac{1}{z_1-H_0}
    \hat{D}
    \frac{1}{z_2-H_0}
    \hat{D}
    \frac{1}{z_1-H_0}
    \hat{D}
    \right\rangle,
    \label{eq:two-phot}
\end{equation}
where $z_1=h\nu_1+E_0+i\eta$ and $z_2=h\nu_1+h\nu_2+E_0+i\eta$. We set $\eta=\SI{10}{\milli\electronvolt}$ for the calculations shown here.

Figure~\ref{fig:thz_spec}(a) shows $G_{\mathrm{2ph}}$ in the $(\nu_1,\nu_2)$ plane. Vertical features mark one-photon transitions into the HO manifold, matching the one-photon absorption features shown in the bottom subpanel. The dominant antidiagonal feature, at fixed $\nu_1+\nu_2\approx\SI{60}{\tera\hertz}$, identifies the resonant two-photon process into the high-pairing HE states. Setting $\nu_1=\nu_2\equiv\nu$ along this antidiagonal recovers the $\nu\approx\SI{30}{\tera\hertz}$ single-color pairing resonance reported above, directly confirming its two-photon origin. Since the intermediate state is highly off-resonant, the low-energy part of the two-photon spectrum can also be approximately reproduced using an effective $D^2$ operator, as shown in Supplemental Material, Sec.~\ref{sec:two_phto_ab_sup}.

Figure~\ref{fig:thz_spec}(b) shows the same spectrum reparametrized as $(\nu_1,\nu_1+\nu_2)$, so that the vertical axis directly tracks the energy of the state reached after two photons. The HE resonance then appears as a horizontal line rather than an antidiagonal. Features below the $\nu_1=\nu_2$ diagonal correspond to a Raman-like process in which stimulated emission occurs from HO into lower-lying even-parity states, rather than to sequential-absorption pathways.

These two representations map onto two complementary experimental protocols. Figure~\ref{fig:thz_spec}(a) corresponds to a two-color pump, where the sample is driven simultaneously at $\nu_1$ and $\nu_2$ and the two frequencies are scanned independently. Figure~\ref{fig:thz_spec}(b) corresponds instead to a time-delay protocol, where Fourier transforming with respect to the time delay between pulses gives direct access to the sum-frequency axis $\nu_1+\nu_2$---and hence to the HE energy---without needing to resolve $\nu_1$ and $\nu_2$ separately. Both protocols can reveal the nonlinear pathway into the high-pairing HE manifold. The two-color pump further suggests that separately tuning the two drive frequencies to satisfy the resonant-sum condition $h\nu_1+h\nu_2=E_{\mathrm{HE}}-E_{\mathrm{GS}}$ may provide more efficient access to the paired state than monochromatic driving.

This mechanism also naturally accounts for the strong nonlinear field dependence of the resonant response. Because the relevant transition is two-photon in character, the induced pairing grows nonlinearly with field amplitude and eventually saturates as the target excited-state manifold becomes substantially populated. This behavior is qualitatively consistent with the strong fluence dependence observed experimentally~\cite{rowe_resonant_2023} and distinguishes the present mechanism from ordinary linear absorption into a metallic continuum.

\section{Discussion and Outlook}

\textit{Physical implications.---}
To conclude, our results have three main implications. First, they provide microscopic support for interpreting the \SI{10}{\tera\hertz} resonance in K$_3$C$_{60}$ in terms of resonantly enhanced pairing correlations rather than simple conductivity enhancement alone. Second, they identify a symmetry-based mechanism that may influence, but cannot by itself determine, the lifetime of the photoinduced state. The high-pairing target state has even parity and is reached through an intermediate odd-parity sector. Consequently, its direct return to the even-parity ground state by emission of a single dipole-active excitation is forbidden in the inversion-symmetric electronic model. This restriction may reduce one particular relaxation channel relative to the decay of an optically bright odd-parity state. It does not, however, imply symmetry protection of the photoinduced state: relaxation can proceed through electron--electron scattering, doublon-holon recombination accompanied by the creation of multiple electronic excitations, and coupling to intramolecular or lattice vibrations. Whether the resonantly accessed state is metastable therefore depends on the competition between these decay channels and possible kinetic or energetic bottlenecks, which are not included in the isolated electronic dynamics studied here. Such bottlenecks could nevertheless contribute to the unusually long-lived response observed under favorable driving conditions~\cite{budden_evidence_2021}.

Third, the state tomography of Fig.~\ref{fig:tomography} carries a conceptual implication. In the thermodynamic limit, the three-orbital model with inverted Hund coupling is an $s$-wave superconductor at equilibrium~\cite{nomura_unified_2015,capone2009}, whereas the half-filled repulsive one-orbital Hubbard model is not. Nevertheless, the two models host nearly identical high-pairing HE states and an almost identical resonance phenomenology. If this similarity persists in the macroscopic limit, it would suggest that the light-induced paired state at elevated temperatures is remarkably insensitive to the nature of the equilibrium ground state. This, in turn, would suggest that light-induced superconductivity is not a simple extension of the equilibrium condensate to higher temperatures, but rather a distinct nonequilibrium state of matter. This viewpoint also rationalizes the observation that photoexcitation of the equilibrium superconductor below $T_c$ suppresses the superconducting response, whereas driving the normal state above $T_c$ induces it~\cite{mitrano_possible_2016}.

More broadly, our results support the idea that light-induced coherence is particularly likely to emerge in materials that already host strong pairing fluctuations or other unconventional correlated normal states above equilibrium $T_c$, as has been discussed experimentally in connection with anomalous Nernst response and precursor superconductivity~\cite{Jotzu2023}. In this sense, the present mechanism may provide a concrete microscopic route by which fluctuating paired states become optically amplified and partially stabilized out of equilibrium. In summary, our results imply a control principle for driven quantum materials: engineering optical access to strongly paired and phase-coherent excited states, rather than perturbing low-energy quasiparticles.

\textit{Future directions.---}
Our results open a number of pressing questions for future work, which we briefly summarize here, moving from microscopic theory toward emergent phenomenology.

(i) First, the \SI{10}{\tera\hertz} resonance in the fulleride should be tested in calculations that more directly approach the thermodynamic limit. Promising next steps include embedding strategies based on multiorbital nonequilibrium dynamical mean-field theory and related methods~\cite{aoki_nonequilibrium_2014}, as well as approaches based on infinite tensor-network methods~\cite{Orus2008}.

(ii) Second, the experimentally observed coincidence---across several material classes---between light-induced superconducting-like states and unconventional normal states above $T_c$~\cite{cavalleri_photo-induced_2018,Uemura2019,Buzzi_2021,Jotzu2023} suggests that light-induced coherent pairing may reflect a more general mechanism. Our results for three- and one-orbital models therefore motivate a systematic search for analogous resonant pathways in intermediate-coupling Hubbard materials, especially near the sweet spot $U\simeq W$, including cuprates, nickelates, organic superconductors such as $\kappa$-(BEDT-TTF)$_2X$, heavy-fermion materials, and possibly pnictides, in suitable parts of their phase diagrams. In such systems, the key question is not only whether paired excited states exist, but also how efficiently they can be reached optically through symmetry-allowed intermediate manifolds. This perspective also sharpens the experimental outlook: because the relevant process is intrinsically sequential and two-photon in character, multidimensional spectroscopy and especially two-color pump protocols tuned separately to the GS$\to$HO and HO$\to$HE resonances should provide a particularly incisive route to testing the mechanism, as outlined by our nonlinear-spectroscopy calculations.

(iii) An important open question concerns the ultimate fate of the high-energy paired state at larger system sizes and longer length scales. Besides an $s$-wave superconducting phase, other possibilities appear equally intriguing. One candidate that has been discussed in the nonequilibrium context is a finite-momentum pair condensate analogous to $\eta$-pairing states. These have been predicted for photo-doped, particle-hole-symmetric Hubbard models on bipartite lattices, where the $\eta$-pairing operators generate an exact symmetry of the Hamiltonian~\cite{kaneko2019,Tindall2019,kaneko2020,Li2020,Ueda2024}. The fcc lattice of K$_3$C$_{60}$ is non-bipartite, so this exact algebraic protection does not apply. However, this does not preclude finite-momentum pairing of this kind, and related instabilities have indeed been found to persist in frustrated lattice geometries~\cite{Tindall2020} and particle-hole-asymmetric settings~\cite{werner2026}. It remains an open question whether the doublon-holon pairs identified here evolve into an analogous finite-momentum condensate in the thermodynamic limit. Working out concrete experimental consequences---particularly for nonlinear optical pathways, collective electrodynamics, magnetic response, and probes sensitive to superconducting fluctuations---should help distinguish between these possibilities.

Finally, coupling the electronic resonance to a dark or driven cavity offers an intriguing route toward modifying superconductivity. Recent reports of cavity-altered superconductivity across diverse materials and platforms~\cite{Keren2026,xu2026,montanaro2026,zhang2026} make this a promising direction for future exploration.

\begin{acknowledgments}
Discussions with Gregor Jotzu at an early stage of the project are gratefully acknowledged. We also acknowledge valuable feedback by Andrea Cavalleri and Rupert Huber on an earlier draft of the manuscript. M.A.S. and P.F. were funded by the European Union (ERC, CAVMAT, project no.~101124492). Views and opinions expressed are, however, those of the author(s) only and do not necessarily reflect those of the European Union or the European Research Council. Neither the European Union nor the European Research Council can be held responsible for them. M.A.S. acknowledges funding by the Deutsche Forschungsgemeinschaft (DFG, German Research Foundation) through project no.~531215165 (Research Unit OPTIMAL). We acknowledge support from the Max Planck--New York City Center for Non-Equilibrium Quantum Phenomena. The Flatiron Institute is a division of the Simons Foundation.
\end{acknowledgments}

\bibliography{k3c60_references1}

@article{cavalleri_photo-induced_2018,
	title = {Photo-induced superconductivity},
	volume = {59},
	issn = {0010-7514},
	url = {https://doi.org/10.1080/00107514.2017.1406623},
	doi = {10.1080/00107514.2017.1406623},
	number = {1},
	urldate = {2019-01-28},
	journal = {Contemporary Physics},
	author = {Cavalleri, Andrea},
	month = jan,
	year = {2018},
	pages = {31--46},
}

@article{fausti_light-induced_2011,
	title = {Light-{Induced} {Superconductivity} in a {Stripe}-{Ordered} {Cuprate}},
	volume = {331},
	issn = {0036-8075, 1095-9203},
	url = {http://www.sciencemag.org/content/331/6014/189},
	doi = {10.1126/science.1197294},
	number = {6014},
	urldate = {2015-03-16},
	journal = {Science},
	author = {Fausti, D. and Tobey, R. I. and Dean, N. and Kaiser, S. and Dienst, A. and Hoffmann, M. C. and Pyon, S. and Takayama, T. and Takagi, H. and Cavalleri, A.},
	month = jan,
	year = {2011},
	pmid = {21233381},
	pages = {189--191},
}

@article{nava_cooling_2018,
	title = {Cooling quasiparticles in {A}$_3${C}$_{60}$ fullerides by excitonic mid-infrared absorption},
	volume = {14},
	copyright = {2017 Nature Publishing Group},
	issn = {1745-2481},
	url = {https://www.nature.com/articles/nphys4288},
	doi = {10.1038/nphys4288},
	number = {2},
	urldate = {2021-03-17},
	journal = {Nature Physics},
	author = {Nava, Andrea and Giannetti, Claudio and Georges, Antoine and Tosatti, Erio and Fabrizio, Michele},
	month = feb,
	year = {2018},
	pages = {154--159},
}

@article{mitrano_possible_2016,
	title = {Possible light-induced superconductivity in {K}$_3${C}$_{60}$ at high temperature},
	volume = {530},
	issn = {1476-4687},
	url = {https://www.nature.com/articles/nature16522},
	doi = {10.1038/nature16522},
	number = {7591},
	urldate = {2025-11-26},
	journal = {Nature},
	author = {Mitrano, M. and Cantaluppi, A. and Nicoletti, D. and Kaiser, S. and Perucchi, A. and Lupi, S. and Di Pietro, P. and Pontiroli, D. and Ricc{\`o}, M. and Clark, S. R. and Jaksch, D. and Cavalleri, A.},
	month = feb,
	year = {2016},
	pages = {461--464},
}

@article{rowe_resonant_2023,
	title = {Resonant enhancement of photo-induced superconductivity in {K}$_3${C}$_{60}$},
	volume = {19},
	issn = {1745-2481},
	url = {https://www.nature.com/articles/s41567-023-02235-9},
	doi = {10.1038/s41567-023-02235-9},
	number = {12},
	urldate = {2025-11-26},
	journal = {Nature Physics},
	author = {Rowe, E. and Yuan, B. and Buzzi, M. and Jotzu, G. and Zhu, Y. and Fechner, M. and F{\"o}rst, M. and Liu, B. and Pontiroli, D. and Ricc{\`o}, M. and Cavalleri, A.},
	month = dec,
	year = {2023},
	pages = {1821--1826},
}

@article{basov_towards_2017,
	title = {Towards properties on demand in quantum materials},
	volume = {16},
	issn = {1476-4660},
	url = {https://www.nature.com/articles/nmat5017},
	doi = {10.1038/nmat5017},
	number = {11},
	urldate = {2025-11-27},
	journal = {Nature Materials},
	author = {Basov, D. N. and Averitt, R. D. and Hsieh, D.},
	month = nov,
	year = {2017},
	pages = {1077--1088},
}

@article{de_la_torre_colloquium_2021,
	title = {Colloquium: {Nonthermal} pathways to ultrafast control in quantum materials},
	volume = {93},
	shorttitle = {Colloquium},
	url = {https://link.aps.org/doi/10.1103/RevModPhys.93.041002},
	doi = {10.1103/RevModPhys.93.041002},
	number = {4},
	urldate = {2025-11-27},
	journal = {Reviews of Modern Physics},
	author = {de la Torre, Alberto and Kennes, Dante M. and Claassen, Martin and Gerber, Simon and McIver, James W. and Sentef, Michael A.},
	month = oct,
	year = {2021},
	pages = {041002},
}

@article{cantaluppi_pressure_2018,
	title = {Pressure tuning of light-induced superconductivity in {K}$_3${C}$_{60}$},
	volume = {14},
	issn = {1745-2481},
	url = {https://www.nature.com/articles/s41567-018-0134-8},
	doi = {10.1038/s41567-018-0134-8},
	number = {8},
	urldate = {2025-11-27},
	journal = {Nature Physics},
	author = {Cantaluppi, A. and Buzzi, M. and Jotzu, G. and Nicoletti, D. and Mitrano, M. and Pontiroli, D. and Ricc{\`o}, M. and Perucchi, A. and Di Pietro, P. and Cavalleri, A.},
	month = aug,
	year = {2018},
	pages = {837--841},
}

@article{budden_evidence_2021,
	title = {Evidence for metastable photo-induced superconductivity in {K}$_3${C}$_{60}$},
	volume = {17},
	issn = {1745-2481},
	url = {https://www.nature.com/articles/s41567-020-01148-1},
	doi = {10.1038/s41567-020-01148-1},
	number = {5},
	urldate = {2025-11-27},
	journal = {Nature Physics},
	author = {Budden, M. and Gebert, T. and Buzzi, M. and Jotzu, G. and Wang, E. and Matsuyama, T. and Meier, G. and Laplace, Y. and Pontiroli, D. and Ricc{\`o}, M. and Schlawin, F. and Jaksch, D. and Cavalleri, A.},
	month = may,
	year = {2021},
	pages = {611--618},
}

@article{buzzi_photomolecular_2020,
	title = {Photomolecular {High}-{Temperature} {Superconductivity}},
	volume = {10},
	url = {https://link.aps.org/doi/10.1103/PhysRevX.10.031028},
	doi = {10.1103/PhysRevX.10.031028},
	number = {3},
	urldate = {2025-11-27},
	journal = {Physical Review X},
	author = {Buzzi, M. and Nicoletti, D. and Fechner, M. and Tancogne-Dejean, N. and Sentef, M. A. and Georges, A. and Biesner, T. and Uykur, E. and Dressel, M. and Henderson, A. and Siegrist, T. and Schlueter, J. A. and Miyagawa, K. and Kanoda, K. and Nam, M.-S. and Ardavan, A. and Coulthard, J. and Tindall, J. and Schlawin, F. and Jaksch, D. and Cavalleri, A.},
	month = aug,
	year = {2020},
	pages = {031028},
}

@article{aoki_nonequilibrium_2014,
	title = {Nonequilibrium dynamical mean-field theory and its applications},
	volume = {86},
	url = {https://link.aps.org/doi/10.1103/RevModPhys.86.779},
	doi = {10.1103/RevModPhys.86.779},
	number = {2},
	urldate = {2025-12-15},
	journal = {Reviews of Modern Physics},
	author = {Aoki, Hideo and Tsuji, Naoto and Eckstein, Martin and Kollar, Marcus and Oka, Takashi and Werner, Philipp},
	month = jun,
	year = {2014},
	pages = {779--837},
}

@article{capone2009,
  title = {Colloquium: {Modeling} the unconventional superconducting properties of expanded {A}$_{3}${C}$_{60}$ fullerides},
  author = {Capone, Massimo and Fabrizio, Michele and Castellani, Claudio and Tosatti, Erio},
  journal = {Rev. Mod. Phys.},
  volume = {81},
  issue = {2},
  pages = {943--958},
  numpages = {0},
  year = {2009},
  month = {Jun},
  publisher = {American Physical Society},
  doi = {10.1103/RevModPhys.81.943},
  url = {https://link.aps.org/doi/10.1103/RevModPhys.81.943}
}

@article{nomura2012_abinitio,
  title = {Ab initio derivation of electronic low-energy models for {C}$_{60}$ and aromatic compounds},
  author = {Nomura, Yusuke and Nakamura, Kazuma and Arita, Ryotaro},
  journal = {Phys. Rev. B},
  volume = {85},
  issue = {15},
  pages = {155452},
  numpages = {12},
  year = {2012},
  month = {Apr},
  publisher = {American Physical Society},
  doi = {10.1103/PhysRevB.85.155452},
  url = {https://link.aps.org/doi/10.1103/PhysRevB.85.155452}
}

@article{Kim_2016,
  title = {Enhancing superconductivity in ${A}_{3}${C}$_{60}$ fullerides},
  author = {Kim, Minjae and Nomura, Yusuke and Ferrero, Michel and Seth, Priyanka and Parcollet, Olivier and Georges, Antoine},
  journal = {Phys. Rev. B},
  volume = {94},
  issue = {15},
  pages = {155152},
  numpages = {12},
  year = {2016},
  month = {Oct},
  publisher = {American Physical Society},
  doi = {10.1103/PhysRevB.94.155152},
  url = {https://link.aps.org/doi/10.1103/PhysRevB.94.155152}
}

@misc{chattopadhyay2026,
      title={Giant Resonant Enhancement of Photoinduced Dynamical Cooper Pairing, far above {T}$_c$}, 
      author={Sambuddha Chattopadhyay and Marios Michael and Andrea Cavalleri and Eugene Demler},
      year={2026},
      eprint={2601.18712},
      archivePrefix={arXiv},
      primaryClass={cond-mat.supr-con},
      url={https://arxiv.org/abs/2601.18712}, 
}

@article{Buzzi_2021,
  title = {Higgs-Mediated Optical Amplification in a Nonequilibrium Superconductor},
  author = {Buzzi, Michele and Jotzu, Gregor and Cavalleri, Andrea and Cirac, J. Ignacio and Demler, Eugene A. and Halperin, Bertrand I. and Lukin, Mikhail D. and Shi, Tao and Wang, Yao and Podolsky, Daniel},
  journal = {Phys. Rev. X},
  volume = {11},
  issue = {1},
  pages = {011055},
  numpages = {16},
  year = {2021},
  month = {Mar},
  publisher = {American Physical Society},
  doi = {10.1103/PhysRevX.11.011055},
  url = {https://link.aps.org/doi/10.1103/PhysRevX.11.011055}
}

@Article{Kennes2017,
author={Kennes, Dante M.
and Wilner, Eli Y.
and Reichman, David R.
and Millis, Andrew J.},
title={Transient superconductivity from electronic squeezing of optically pumped phonons},
journal={Nature Physics},
year={2017},
month={May},
day={01},
volume={13},
number={5},
pages={479-483},
issn={1745-2481},
doi={10.1038/nphys4024},
url={https://doi.org/10.1038/nphys4024}
}

@article{Knap_2016,
  title = {Dynamical Cooper pairing in nonequilibrium electron-phonon systems},
  author = {Knap, Michael and Babadi, Mehrtash and Refael, Gil and Martin, Ivar and Demler, Eugene},
  journal = {Phys. Rev. B},
  volume = {94},
  issue = {21},
  pages = {214504},
  numpages = {13},
  year = {2016},
  month = {Dec},
  publisher = {American Physical Society},
  doi = {10.1103/PhysRevB.94.214504},
  url = {https://link.aps.org/doi/10.1103/PhysRevB.94.214504}
}

@article{Murakami_2017,
  title = {Nonequilibrium steady states and transient dynamics of conventional superconductors under phonon driving},
  author = {Murakami, Yuta and Tsuji, Naoto and Eckstein, Martin and Werner, Philipp},
  journal = {Phys. Rev. B},
  volume = {96},
  issue = {4},
  pages = {045125},
  numpages = {19},
  year = {2017},
  month = {Jul},
  publisher = {American Physical Society},
  doi = {10.1103/PhysRevB.96.045125},
  url = {https://link.aps.org/doi/10.1103/PhysRevB.96.045125}
}

@article{Babadi_2017,
  title = {Theory of parametrically amplified electron-phonon superconductivity},
  author = {Babadi, Mehrtash and Knap, Michael and Martin, Ivar and Refael, Gil and Demler, Eugene},
  journal = {Phys. Rev. B},
  volume = {96},
  issue = {1},
  pages = {014512},
  numpages = {38},
  year = {2017},
  month = {Jul},
  publisher = {American Physical Society},
  doi = {10.1103/PhysRevB.96.014512},
  url = {https://link.aps.org/doi/10.1103/PhysRevB.96.014512}
}

@article{Mazza_2017,
  title = {Nonequilibrium superconductivity in driven alkali-doped fullerides},
  author = {Mazza, Giacomo and Georges, Antoine},
  journal = {Phys. Rev. B},
  volume = {96},
  issue = {6},
  pages = {064515},
  numpages = {10},
  year = {2017},
  month = {Aug},
  publisher = {American Physical Society},
  doi = {10.1103/PhysRevB.96.064515},
  url = {https://link.aps.org/doi/10.1103/PhysRevB.96.064515}
}

@article{Dasari_2018,
  title = {Transient {Floquet} engineering of superconductivity},
  author = {Dasari, Nagamalleswararao and Eckstein, Martin},
  journal = {Phys. Rev. B},
  volume = {98},
  issue = {23},
  pages = {235149},
  numpages = {10},
  year = {2018},
  month = {Dec},
  publisher = {American Physical Society},
  doi = {10.1103/PhysRevB.98.235149},
  url = {https://link.aps.org/doi/10.1103/PhysRevB.98.235149}
}

@article{Grankin_2026,
  title = {Integrable-to-thermalizing crossover in nonequilibrium superconductors},
  author = {Grankin, Andrey and Galitski, Victor},
  journal = {Phys. Rev. B},
  volume = {113},
  issue = {10},
  pages = {104509},
  numpages = {6},
  year = {2026},
  month = {Mar},
  publisher = {American Physical Society},
  doi = {10.1103/nzyh-hgpr},
  url = {https://link.aps.org/doi/10.1103/nzyh-hgpr}
}

@Article{Chattopadhyay2025,
author={Chattopadhyay, Sambuddha
and Eckhardt, Christian J.
and Kennes, Dante M.
and Sentef, Michael A.
and Shin, Dongbin
and Rubio, Angel
and Cavalleri, Andrea
and Demler, Eugene A.
and Michael, Marios H.},
title={Metastable photo-induced superconductivity far above {T}$_c$},
journal={npj Quantum Materials},
year={2025},
month={Mar},
day={28},
volume={10},
number={1},
pages={34},
issn={2397-4648},
doi={10.1038/s41535-025-00750-x},
url={https://doi.org/10.1038/s41535-025-00750-x}
}

@article{itensor,
	title={{The ITensor Software Library for Tensor Network Calculations}},
	author={Matthew Fishman and Steven R. White and E. Miles Stoudenmire},
	journal={SciPost Phys. Codebases},
	pages={4},
	year={2022},
	publisher={SciPost},
	doi={10.21468/SciPostPhysCodeb.4},
	url={https://scipost.org/10.21468/SciPostPhysCodeb.4}
}

@article{nomura_unified_2015,
	title = {Unified understanding of superconductivity and {Mott} transition in alkali-doped fullerides from first principles},
	volume = {1},
	url = {https://www.science.org/doi/10.1126/sciadv.1500568},
	doi = {10.1126/sciadv.1500568},
	number = {7},
	urldate = {2024-06-14},
	journal = {Science Advances},
	author = {Nomura, Yusuke and Sakai, Shiro and Capone, Massimo and Arita, Ryotaro},
	month = aug,
	year = {2015},
	pages = {e1500568},
}

@article{dargel_2012,
  title = {Lanczos algorithm with matrix product states for dynamical correlation functions},
  author = {Dargel, P. E. and W\"ollert, A. and Honecker, A. and McCulloch, I. P. and Schollw\"ock, U. and Pruschke, T.},
  journal = {Phys. Rev. B},
  volume = {85},
  issue = {20},
  pages = {205119},
  numpages = {11},
  year = {2012},
  month = {May},
  publisher = {American Physical Society},
  doi = {10.1103/PhysRevB.85.205119},
  url = {https://link.aps.org/doi/10.1103/PhysRevB.85.205119}
}

@article{nomura_2015_abinitio,
  title = {Ab initio downfolding for electron-phonon-coupled systems: Constrained density-functional perturbation theory},
  author = {Nomura, Yusuke and Arita, Ryotaro},
  journal = {Phys. Rev. B},
  volume = {92},
  issue = {24},
  pages = {245108},
  numpages = {15},
  year = {2015},
  month = {Dec},
  publisher = {American Physical Society},
  doi = {10.1103/PhysRevB.92.245108},
  url = {https://link.aps.org/doi/10.1103/PhysRevB.92.245108}
}

@article{Jotzu2023,
  title = {Superconducting Fluctuations Observed Far above {T}$_{c}$ in the Isotropic Superconductor {K}$_{3}${C}$_{60}$},
  author = {Jotzu, Gregor and Meier, Guido and Cantaluppi, Alice and Cavalleri, Andrea and Pontiroli, Daniele and Ricc\`o, Mauro and Ardavan, Arzhang and Nam, Moon-Sun},
  journal = {Phys. Rev. X},
  volume = {13},
  issue = {2},
  pages = {021008},
  numpages = {13},
  year = {2023},
  month = {Apr},
  publisher = {American Physical Society},
  doi = {10.1103/PhysRevX.13.021008},
  url = {https://link.aps.org/doi/10.1103/PhysRevX.13.021008}
}

@misc{sous2025,
      title={Ultrafast electronic coherence from slow phonons}, 
      author={Mattia Moroder and Sebastian Paeckel and Matteo Mitrano and John Sous},
      year={2025},
      eprint={2509.06939},
      archivePrefix={arXiv},
      primaryClass={cond-mat.supr-con},
      url={https://arxiv.org/abs/2509.06939}, 
}

@article{Keren2026,
  author = {Keren, Itai and Webb, Tatiana A. and Zhang, Shuai and Xu, Jikai and Sun, Dihao and Kim, Brian S. Y. and Shin, Dongbin and Zhang, Songtian S. and Zhang, Junhe and Pereira, Giancarlo and Yao, Juntao and Okugawa, Takuya and Michael, Marios H. and Viñas Boström, Emil and Edgar, James H. and Wolf, Stuart and Julian, Matthew and Prasankumar, Rohit P. and Miyagawa, Kazuya and Kanoda, Kazushi and Gu, Genda and Cothrine, Matthew and Mandrus, David and Buzzi, Michele and Cavalleri, Andrea and Dean, Cory R. and Kennes, Dante M. and Millis, Andrew J. and Li, Qiang and Sentef, Michael A. and Rubio, Angel and Pasupathy, Abhay N. and Basov, D. N.},
  title = {Cavity-altered superconductivity},
  journal = {Nature},
  year = {2026},
  volume = {650},
  number = {8103},
  pages = {864--868},
  doi = {10.1038/s41586-025-10062-6},
  url = {https://doi.org/10.1038/s41586-025-10062-6},
  isbn = {1476-4687}
}

@misc{montanaro2026,
      title={Cavity-enhanced superconducting response in an underdoped cuprate}, 
      author={Angela Montanaro and Vadim Plastovets and Nitesh Khatiwada and Jacopo Fiore and Giacomo Jarc and Abdullah Alabbadi and Antonio Mastropasqua and Enrico Maria Rigoni and Shahla Y. Mathengattil and Simone Dal Zilio and Francesca Fassioli Olsen and Fabio Novelli and Stephan Winnerl and Michael A. Sentef and Dante M. Kennes and Andrew J. Millis and Francesco Piazza and Daniele Fausti},
      year={2026},
      eprint={2606.18084},
      archivePrefix={arXiv},
      primaryClass={cond-mat.supr-con},
      url={https://arxiv.org/abs/2606.18084}, 
}

@misc{zhang2026,
      title={Cavity Enhanced Superconductivity}, 
      author={Hanxiang Zhang and Zexin Feng and I-Te Lu and Zhiwei Li and Songhao Guo and Qiuyu Shang and Dening Luan and Mingcheng Panmai and Kenji Watanabe and Takashi Taniguchi and Angel Rubio and Weibo Gao},
      year={2026},
      eprint={2606.19171},
      archivePrefix={arXiv},
      primaryClass={cond-mat.supr-con},
      url={https://arxiv.org/abs/2606.19171}, 
}

@article{kaneko2019,
  title = {Photoinduced $\ensuremath{\eta}$ Pairing in the {Hubbard} Model},
  author = {Kaneko, Tatsuya and Shirakawa, Tomonori and Sorella, Sandro and Yunoki, Seiji},
  journal = {Phys. Rev. Lett.},
  volume = {122},
  issue = {7},
  pages = {077002},
  numpages = {6},
  year = {2019},
  month = {Feb},
  publisher = {American Physical Society},
  doi = {10.1103/PhysRevLett.122.077002},
  url = {https://link.aps.org/doi/10.1103/PhysRevLett.122.077002}
}

@article{kaneko2020,
  title = {Charge stiffness and long-range correlation in the optically induced $\ensuremath{\eta}$-pairing state of the one-dimensional Hubbard model},
  author = {Kaneko, Tatsuya and Yunoki, Seiji and Millis, Andrew J.},
  journal = {Phys. Rev. Res.},
  volume = {2},
  issue = {3},
  pages = {032027(R)},
  numpages = {5},
  year = {2020},
  month = {Jul},
  publisher = {American Physical Society},
  doi = {10.1103/PhysRevResearch.2.032027},
  url = {https://link.aps.org/doi/10.1103/PhysRevResearch.2.032027}
}

@article{Ueda2024,
  title = {Photoinduced $\ensuremath{\eta}$-pairing correlation in the Hubbard ladder},
  author = {Ueda, Ryota and Kuroki, Kazuhiko and Kaneko, Tatsuya},
  journal = {Phys. Rev. B},
  volume = {109},
  issue = {7},
  pages = {075122},
  numpages = {7},
  year = {2024},
  month = {Feb},
  publisher = {American Physical Society},
  doi = {10.1103/PhysRevB.109.075122},
  url = {https://link.aps.org/doi/10.1103/PhysRevB.109.075122}
}

@article{Tindall2020,
  title = {Dynamical Order and Superconductivity in a Frustrated Many-Body System},
  author = {Tindall, J. and Schlawin, F. and Buzzi, M. and Nicoletti, D. and Coulthard, J. R. and Gao, H. and Cavalleri, A. and Sentef, M. A. and Jaksch, D.},
  journal = {Phys. Rev. Lett.},
  volume = {125},
  issue = {13},
  pages = {137001},
  numpages = {7},
  year = {2020},
  month = {Sep},
  publisher = {American Physical Society},
  doi = {10.1103/PhysRevLett.125.137001},
  url = {https://link.aps.org/doi/10.1103/PhysRevLett.125.137001}
}

@misc{werner2026,
      title={$\eta$-pairing in metallic and particle-hole asymmetric systems}, 
      author={Philipp Werner and Aaram J. Kim and Lei Geng},
      year={2026},
      eprint={2606.05092},
      archivePrefix={arXiv},
      primaryClass={cond-mat.str-el},
      url={https://arxiv.org/abs/2606.05092}, 
}

@article{Li2020,
  title = {$\ensuremath{\eta}$-paired superconducting hidden phase in photodoped Mott insulators},
  author = {Li, Jiajun and Golez, Denis and Werner, Philipp and Eckstein, Martin},
  journal = {Phys. Rev. B},
  volume = {102},
  issue = {16},
  pages = {165136},
  numpages = {10},
  year = {2020},
  month = {Oct},
  publisher = {American Physical Society},
  doi = {10.1103/PhysRevB.102.165136},
  url = {https://link.aps.org/doi/10.1103/PhysRevB.102.165136}
}

@article{Tindall2019,
  title = {Heating-Induced Long-Range $\ensuremath{\eta}$ Pairing in the Hubbard Model},
  author = {Tindall, J. and Bu\ifmmode \check{c}\else \v{c}\fi{}a, B. and Coulthard, J. R. and Jaksch, D.},
  journal = {Phys. Rev. Lett.},
  volume = {123},
  issue = {3},
  pages = {030603},
  numpages = {7},
  year = {2019},
  month = {Jul},
  publisher = {American Physical Society},
  doi = {10.1103/PhysRevLett.123.030603},
  url = {https://link.aps.org/doi/10.1103/PhysRevLett.123.030603}
}

@inproceedings{shimano2014,
author = {R. Shimano and R. Matsunaga and Y. I. Hamada and A. Sugioka and H. Fujita and K. Makise and Y. Uzawa and H. Terai and Z. Wang and N. Tsuji and H. Aoki},
booktitle = {19th International Conference on Ultrafast Phenomena},
journal = {19th International Conference on Ultrafast Phenomena},
pages = {10.Thu.A.1},
publisher = {Optica Publishing Group},
title = {Higgs Mode and Terahertz Nonlinear Optics in Superconductors},
year = {2014},
url = {https://opg.optica.org/abstract.cfm?URI=UP-2014-10.Thu.A.1},
doi = {10.1364/UP.2014.10.Thu.A.1},
}

@article{liu2025,
author={Liu, Albert},
title={Multidimensional terahertz probes of quantum materials},
journal={npj Quantum Materials},
year={2025},
month={Feb},
day={10},
volume={10},
number={1},
pages={18},
issn={2397-4648},
doi={10.1038/s41535-025-00741-y},
url={https://doi.org/10.1038/s41535-025-00741-y}
}

@article{katsumi2024,
  title = {Revealing Novel Aspects of Light-Matter Coupling by Terahertz Two-Dimensional Coherent Spectroscopy: The Case of the Amplitude Mode in Superconductors},
  author = {Katsumi, Kota and Fiore, Jacopo and Udina, Mattia and Romero, Ralph and Barbalas, David and Jesudasan, John and Raychaudhuri, Pratap and Seibold, Goetz and Benfatto, Lara and Armitage, N. P.},
  journal = {Phys. Rev. Lett.},
  volume = {132},
  issue = {25},
  pages = {256903},
  numpages = {7},
  year = {2024},
  month = {Jun},
  publisher = {American Physical Society},
  doi = {10.1103/PhysRevLett.132.256903},
  url = {https://link.aps.org/doi/10.1103/PhysRevLett.132.256903}
}

@article{Orus2008,
  title = {Infinite time-evolving block decimation algorithm beyond unitary evolution},
  author = {Or\'us, R. and Vidal, G.},
  journal = {Phys. Rev. B},
  volume = {78},
  issue = {15},
  pages = {155117},
  numpages = {11},
  year = {2008},
  month = {Oct},
  publisher = {American Physical Society},
  doi = {10.1103/PhysRevB.78.155117},
  url = {https://link.aps.org/doi/10.1103/PhysRevB.78.155117}
}

@misc{xu2026,
      title={Vacuum-dressed superconductivity in {NbN} observed in a high-{$Q$} terahertz cavity}, 
      author={Hongjing Xu and Andrey Baydin and Qinyan Yi and I-Te Lu and Ningxu Zhu and T. Elijah Kritzell and Jacques Doumani and Dasom Kim and Fuyang Tay and Angel Rubio and Junichiro Kono},
      year={2026},
      eprint={2601.08191},
      archivePrefix={arXiv},
      primaryClass={physics.optics},
      url={https://arxiv.org/abs/2601.08191}, 
}

@article{Uemura2019,
  title = {Dynamic superconductivity responses in photoexcited optical conductivity and {Nernst} effect},
  author = {Uemura, Yasutomo J.},
  journal = {Phys. Rev. Mater.},
  volume = {3},
  issue = {10},
  pages = {104801},
  numpages = {11},
  year = {2019},
  month = {Oct},
  publisher = {American Physical Society},
  doi = {10.1103/PhysRevMaterials.3.104801},
  url = {https://link.aps.org/doi/10.1103/PhysRevMaterials.3.104801}
}

@article{hu_optically_2014,
  title = {Optically enhanced coherent transport in {YBa}$_2${Cu}$_3${O}$_{6.5}$ by ultrafast redistribution of interlayer coupling},
  author = {Hu, W. and Kaiser, S. and Nicoletti, D. and Hunt, C. R. and Gierz, I. and Hoffmann, M. C. and Le Tacon, M. and Loew, T. and Keimer, B. and Cavalleri, A.},
  journal = {Nature Materials},
  volume = {13},
  number = {7},
  pages = {705--711},
  year = {2014},
  month = jul,
  doi = {10.1038/nmat3963},
  url = {https://www.nature.com/articles/nmat3963}
}

@article{kaiser_optically_2014,
  title = {Optically induced coherent transport far above {$T_c$} in underdoped {YBa}$_2${Cu}$_3${O}$_{6+\delta}$},
  author = {Kaiser, S. and Hunt, C. R. and Nicoletti, D. and Hu, W. and Gierz, I. and Liu, H. Y. and Le Tacon, M. and Loew, T. and Haug, D. and Keimer, B. and Cavalleri, A.},
  journal = {Phys. Rev. B},
  volume = {89},
  issue = {18},
  pages = {184516},
  year = {2014},
  month = {May},
  publisher = {American Physical Society},
  doi = {10.1103/PhysRevB.89.184516},
  url = {https://link.aps.org/doi/10.1103/PhysRevB.89.184516}
}

@article{mankowsky_nonlinear_2014,
  title = {Nonlinear lattice dynamics as a basis for enhanced superconductivity in {YBa}$_2${Cu}$_3${O}$_{6.5}$},
  author = {Mankowsky, R. and Subedi, A. and F{\"o}rst, M. and Mariager, S. O. and Chollet, M. and Lemke, H. T. and Robinson, J. S. and Glownia, J. M. and Minitti, M. P. and Frano, A. and Fechner, M. and Spaldin, N. A. and Loew, T. and Keimer, B. and Georges, A. and Cavalleri, A.},
  journal = {Nature},
  volume = {516},
  number = {7529},
  pages = {71--73},
  year = {2014},
  month = dec,
  doi = {10.1038/nature13875},
  url = {https://www.nature.com/articles/nature13875}
}

@article{hunt_dynamical_2016,
  title = {Dynamical decoherence of the light induced interlayer coupling in {YBa}$_2${Cu}$_3${O}$_{6+\delta}$},
  author = {Hunt, C. R. and Nicoletti, D. and Kaiser, S. and Pr{\"o}pper, D. and Loew, T. and Porras, J. and Keimer, B. and Cavalleri, A.},
  journal = {Phys. Rev. B},
  volume = {94},
  issue = {22},
  pages = {224303},
  year = {2016},
  month = {Dec},
  publisher = {American Physical Society},
  doi = {10.1103/PhysRevB.94.224303},
  url = {https://link.aps.org/doi/10.1103/PhysRevB.94.224303}
}

@article{buzzi_phase_2021,
  title = {Phase Diagram for Light-Induced Superconductivity in $\kappa$-({ET})$_2$-{X}},
  author = {Buzzi, M. and Nicoletti, D. and Fava, S. and Jotzu, G. and Miyagawa, K. and Kanoda, K. and Henderson, A. and Siegrist, T. and Schlueter, J. A. and Nam, M.-S. and Ardavan, A. and Cavalleri, A.},
  journal = {Phys. Rev. Lett.},
  volume = {127},
  issue = {19},
  pages = {197002},
  year = {2021},
  month = {Nov},
  publisher = {American Physical Society},
  doi = {10.1103/PhysRevLett.127.197002},
  url = {https://link.aps.org/doi/10.1103/PhysRevLett.127.197002}
}

@article{fava_magnetic_2024,
  title = {Magnetic field expulsion in optically driven {YBa}$_2${Cu}$_3${O}$_{6.48}$},
  author = {Fava, S. and De Vecchi, G. and Jotzu, G. and Buzzi, M. and Gebert, T. and Liu, Y. and Keimer, B. and Cavalleri, A.},
  journal = {Nature},
  volume = {632},
  number = {8023},
  pages = {75--80},
  year = {2024},
  month = aug,
  doi = {10.1038/s41586-024-07635-2},
  url = {https://www.nature.com/articles/s41586-024-07635-2}
}


\clearpage
\onecolumngrid

\setcounter{equation}{0}
\setcounter{figure}{0}
\setcounter{table}{0}
\setcounter{section}{0}

\renewcommand{\theequation}{S\arabic{equation}}
\renewcommand{\thefigure}{S\arabic{figure}}
\renewcommand{\thetable}{S\arabic{table}}
\renewcommand{\thesection}{S\arabic{section}}

\renewcommand{\theHequation}{S\arabic{equation}}
\renewcommand{\theHfigure}{S\arabic{figure}}
\renewcommand{\theHtable}{S\arabic{table}}
\renewcommand{\theHsection}{S\arabic{section}}

\begin{center}
{\large\textbf{Supplemental Material for}}\\[3mm]
{\Large\textbf{Microscopic mechanism for resonant light-enhanced pair correlations in K$_3$C$_{60}$}}
\end{center}

\section{Model Hamiltonian}
\label{sec:sup_hamiltonian}

K$_3$C$_{60}$ has three half-filled $t_{1u}$ states at the Fermi energy that are well isolated from other states. It has been shown that a minimal model containing these states, coupled to Jahn--Teller phonons, describes the equilibrium superconductivity~\cite{nomura_unified_2015}. The minimal electronic model is described by a Hubbard--Kanamori Hamiltonian in which the Jahn--Teller phonons lead to an inverted Hund coupling. The interaction term decomposes into density--density, spin-flip, and pair-hopping contributions, $H_{\mathrm{int}}=H_{\mathrm{nn}}+H_{\mathrm{sf}}+H_{\mathrm{ph}}$.

\begin{equation}
    \begin{split}
        H_{\mathrm{nn}}={}&
        U\sum_\alpha
        \hat{n}_{\alpha,\uparrow}
        \hat{n}_{\alpha,\downarrow}
        \\
        &+
        \left(U-2J_{\mathrm{inv}}\right)
        \sum_{\alpha\neq\beta}
        \hat{n}_{\alpha,\uparrow}
        \hat{n}_{\beta,\downarrow}
        \\
        &+
        \left(U-3J_{\mathrm{inv}}\right)
        \sum_{\alpha<\beta,\sigma}
        \hat{n}_{\alpha,\sigma}
        \hat{n}_{\beta,\sigma},
    \end{split}
    \nonumber
\end{equation}
\begin{equation}
    \begin{split}
        H_{\mathrm{sf}}
        =
        -J_{\mathrm{inv}}
        \sum_{\alpha\neq\beta}
        c^\dagger_{\alpha,\uparrow}
        c^{\phantom{\dagger}}_{\alpha,\downarrow}
        c^\dagger_{\beta,\downarrow}
        c^{\phantom{\dagger}}_{\beta,\uparrow},
    \end{split}
    \nonumber
\end{equation}
\begin{equation}
    \begin{split}
        H_{\mathrm{ph}}
        =
        J_{\mathrm{inv}}
        \sum_{\alpha\neq\beta}
        c^\dagger_{\alpha,\uparrow}
        c^\dagger_{\alpha,\downarrow}
        c^{\phantom{\dagger}}_{\beta,\downarrow}
        c^{\phantom{\dagger}}_{\beta,\uparrow}.
    \end{split}
\end{equation}

For all calculations presented here, we use $U=\SI{500}{\milli\electronvolt}\simeq W$ and $J_{\mathrm{inv}}=\SI{-20}{\milli\electronvolt}$~\cite{nomura2012_abinitio,nomura_2015_abinitio}. We use an \textit{ab initio}-derived nearest-neighbor (NN) and next-nearest-neighbor (NNN) model~\cite{nomura2012_abinitio}, constructed from the Wannierized three $t_{1u}$ bands at the Fermi level:
\begin{equation}
    \begin{split}
        H_{\mathrm{hop}}
        =
        \sum_{i\neq j}
        \sum_{\alpha,\beta,\sigma}
        T_{\alpha,\beta}
        (\mathbf{R}_i-\mathbf{R}_j)
        \left(
        c^\dagger_{i,\alpha,\sigma}
        c^{\phantom{\dagger}}_{j,\beta,\sigma}
        +\mathrm{h.c.}
        \right),
    \end{split}
    \label{eq:sup-hubb-kana}
\end{equation}
where $T_{\alpha,\beta}(\mathbf{R}_i-\mathbf{R}_j)$ is the hopping matrix as a function of the hopping vector. The NN hopping matrices are
\begin{align*}
\begin{pmatrix}
F_1 & F_2 & 0\\
F_2 & F_3 & 0\\
0 & 0 & F_4
\end{pmatrix}
&\quad\text{for}\quad
\mathbf{R}=(0.5,0.5,0.0),
\\[8pt]
\begin{pmatrix}
F_4 & 0 & 0\\
0 & F_1 & F_2\\
0 & F_2 & F_3
\end{pmatrix}
&\quad\text{for}\quad
\mathbf{R}=(0.0,0.5,0.5),
\\[8pt]
\begin{pmatrix}
F_4 & 0 & 0\\
0 & F_1 & -F_2\\
0 & -F_2 & F_3
\end{pmatrix}
&\quad\text{for}\quad
\mathbf{R}=(0.0,0.5,-0.5),
\\[12pt]
\begin{pmatrix}
F_3 & 0 & F_2\\
0 & F_4 & 0\\
F_2 & 0 & F_1
\end{pmatrix}
&\quad\text{for}\quad
\mathbf{R}=(0.5,0.0,0.5),
\\[8pt]
\begin{pmatrix}
F_3 & 0 & -F_2\\
0 & F_4 & 0\\
-F_2 & 0 & F_1
\end{pmatrix}
&\quad\text{for}\quad
\mathbf{R}=(-0.5,0.0,0.5),
\\[12pt]
\begin{pmatrix}
F_1 & -F_2 & 0\\
-F_2 & F_3 & 0\\
0 & 0 & F_4
\end{pmatrix}
&\quad\text{for}\quad
\mathbf{R}=(0.5,-0.5,0.0).
\end{align*}

Here $\mathbf{R}=\pm(\mathbf{R}_i-\mathbf{R}_j)$ is expressed in units of the lattice constant $a_{\mathrm{lattice}}=\SI{14.24}{\angstrom}$. The NNN hopping matrices are
\begin{align*}
\begin{pmatrix}
F_5 & 0 & 0\\
0 & F_6 & 0\\
0 & 0 & F_7
\end{pmatrix}
&\quad\text{for}\quad
\mathbf{R}=(1,0,0),
\\[8pt]
\begin{pmatrix}
F_7 & 0 & 0\\
0 & F_5 & 0\\
0 & 0 & F_6
\end{pmatrix}
&\quad\text{for}\quad
\mathbf{R}=(0,1,0),
\\[8pt]
\begin{pmatrix}
F_6 & 0 & 0\\
0 & F_7 & 0\\
0 & 0 & F_5
\end{pmatrix}
&\quad\text{for}\quad
\mathbf{R}=(0,0,1).
\end{align*}
The hopping values are $F_1=\SI{-4}{\milli\electronvolt}$, $F_2=\SI{-33.9}{\milli\electronvolt}$, $F_3=\SI{42.1}{\milli\electronvolt}$, $F_4=\SI{-18.7}{\milli\electronvolt}$, $F_5=\SI{-9.4}{\milli\electronvolt}$, $F_6=\SI{-1.4}{\milli\electronvolt}$, and $F_7=\SI{-0.2}{\milli\electronvolt}$.

Finally, we describe the coupling between the drive and the electrons in the length gauge:
\begin{equation}
    \begin{split}
        H_{\mathrm{drive}}(t)
        &=
        E\sin(2\pi\nu t)D^x
        \\
        &=
        E\sin(2\pi\nu t)
        \sum_{i,\alpha}
        R_i^x\hat{n}_{i,\alpha},
    \end{split}
\end{equation}
where $E$ is the electric-field amplitude and $R_i^x$ is the $x$ component of the position of the $i$th site of the fcc cluster. We also checked our results by coupling the field via the Peierls substitution in the two-site cluster and obtained results identical to those produced using $H_{\mathrm{drive}}(t)$. We adopt the length-gauge representation because it is computationally more efficient. Adding a Gaussian envelope to the electric-field pulse also does not change the results in a meaningful way. The total time-dependent Hamiltonian is
\begin{equation}
    H_{\mathrm{tot}}
    =
    H_{\mathrm{nn}}
    +H_{\mathrm{sf}}
    +H_{\mathrm{ph}}
    +H_{\mathrm{hop}}
    +H_{\mathrm{drive}}(t).
\end{equation}

\section{Drive-induced couplings}
\label{sec:sup_drivesym}

To understand the driven Hamiltonian, it is useful to examine how the drive couples different symmetry sectors. The K$_3$C$_{60}$ fcc lattice has inversion symmetry; consequently, the Hubbard--Kanamori Hamiltonian describing it also respects this symmetry. The parity operator $\Pi$ commutes with the equilibrium Hamiltonian, $[H_0,\Pi]=0$, and satisfies $\Pi^2=1$. Its eigenvalues are therefore $\pm1$, and the eigenstates of $H_0$ can be classified as having even or odd parity. For the parameters describing K$_3$C$_{60}$, we find that the ground state is even under inversion:
\begin{equation}
    \Pi|\mathrm{GS}\rangle
    =
    +|\mathrm{GS}\rangle.
\end{equation}

The drive term is proportional to the dipole operator
\begin{equation}
    D^x
    =
    \sum_{i,\alpha}
    R_i^x\hat{n}_{i,\alpha}.
\end{equation}
This operator is odd under inversion:
\begin{equation}
    \Pi^{-1}D^x\Pi=-D^x.
\end{equation}
Equivalently,
\begin{equation}
    [D^x,\Pi]=-2\Pi D^x.
\end{equation}

Suppose that $|E_1\rangle$ and $|E_2\rangle$ are both even-parity states. Their dipole matrix element satisfies
\begin{equation}
    \begin{split}
        \langle E_1|D^x|E_2\rangle
        &=
        \langle E_1|
        \Pi^{-1}D^x\Pi
        |E_2\rangle
        \\
        &=
        -\langle E_1|D^x|E_2\rangle,
    \end{split}
\end{equation}
which implies
\begin{equation}
    \langle E_1|D^x|E_2\rangle=0.
\end{equation}
The same argument applies when both states have odd parity. The drive therefore couples only states of opposite parity.

The driven Hamiltonian can be represented in the equilibrium eigenbasis as
\begin{equation}
H(t)=
\left(
\begin{array}{@{}c|c|c@{}}
E_{\mathrm{GS}}
&
V_{\mathrm{GS,O}}(t)
&
0
\\ \hline
V_{\mathrm{GS,O}}^\dagger(t)
&
H_{\mathrm{O}}
&
V_{\mathrm{O,E}}(t)
\\ \hline
0
&
V_{\mathrm{O,E}}^\dagger(t)
&
H_{\mathrm{E}}
\end{array}
\right),
\end{equation}
where $H_{\mathrm{O}}$ and $H_{\mathrm{E}}$ are diagonal blocks containing the odd- and even-parity excited states, respectively. The drive blocks $V_{\mathrm{GS,O}}(t)$ and $V_{\mathrm{O,E}}(t)$ couple only opposite-parity sectors.

\section{Exact diagonalization on the three-orbital dimer}
\label{sec:two-bucky-ED}

\begin{figure}[h]
    \centering
    \includegraphics[width=0.9\textwidth]{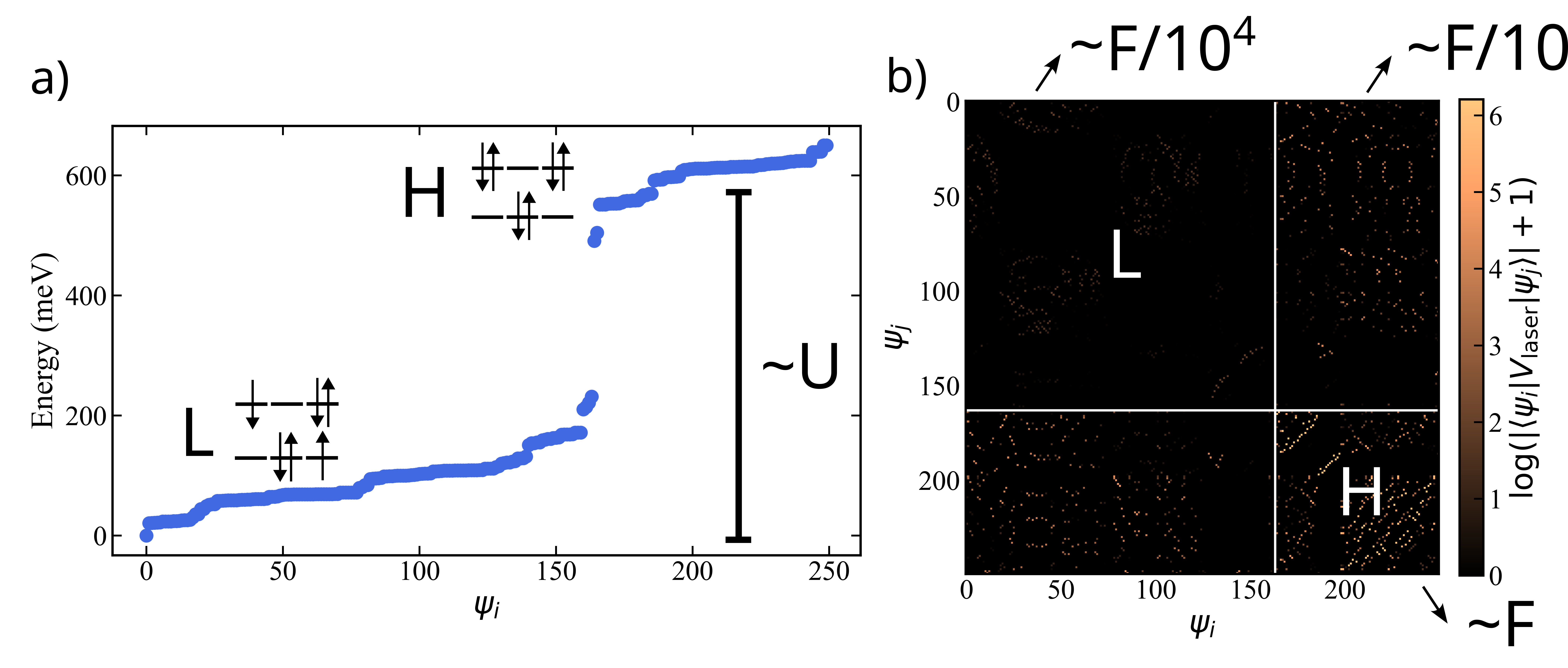}
    \caption{
    \textbf{Spectrum and dipole-coupling hierarchy of the two-buckyball system.}
    \textbf{(a)} Eigenenergy spectrum, separating low- (L) and high-energy (H) sectors according to doublon localization.
    \textbf{(b)} Drive term expressed in the eigenbasis of the undriven Hamiltonian, showing the hierarchy of dipole couplings.
    }
    \label{fig:twobuckyball}
\end{figure}

Here we study the two-buckyball system, which is the minimal three-orbital system that reproduces the resonances in the pair correlation while keeping the Hilbert-space dimension small enough to compute all relevant quantities by exact diagonalization.

We examine the spectrum of the equilibrium Hamiltonian and the matrix elements of the driving term expressed in the equilibrium eigenbasis, shown in Figs.~\ref{fig:twobuckyball}(a) and~\ref{fig:twobuckyball}(b), respectively. In panel (a), we distinguish a low-energy sector (L) and a high-energy sector (H). In the former, most of the weight of the eigenstates corresponds to three electrons per buckyball, $(3,3)$, while in the latter the dominant components correspond to four electrons on one buckyball and two on the other, $(4,2)$. This distinction determines how the drive acts. Since the drive is proportional to the dipole operator, states in which the buckyballs do not strongly deviate from a $(3,3)$ electronic distribution have suppressed couplings of order $F/10^4$, where $F$ denotes the coupling strength of the drive to the material. In the H sector, by contrast, most of the weight lies in the $(4,2)$ electronic distribution, leading to large dipole moments and strong couplings of order $F$. The coupling between the L and H sectors has an intermediate strength of order $F/10$.

Using these symmetry constraints and coupling hierarchies, we construct an effective Hamiltonian in the reduced basis
\begin{equation}
    B_{\mathrm{eff}}
    =
    \{
    |\mathrm{GS}\rangle;
    |\mathrm{HO}_0\rangle,\ldots,|\mathrm{HO}_{N_{\mathrm{HO}}}\rangle;
    |\mathrm{HE}_0\rangle,\ldots,|\mathrm{HE}_{N_{\mathrm{HE}}}\rangle
    \}.
\end{equation}
Here, $|\mathrm{HO}_i\rangle$ and $|\mathrm{HE}_i\rangle$ are states in the high-energy H sector with odd and even inversion parity, respectively.

\begin{figure}[h]
    \centering
    \includegraphics[width=0.9\textwidth]{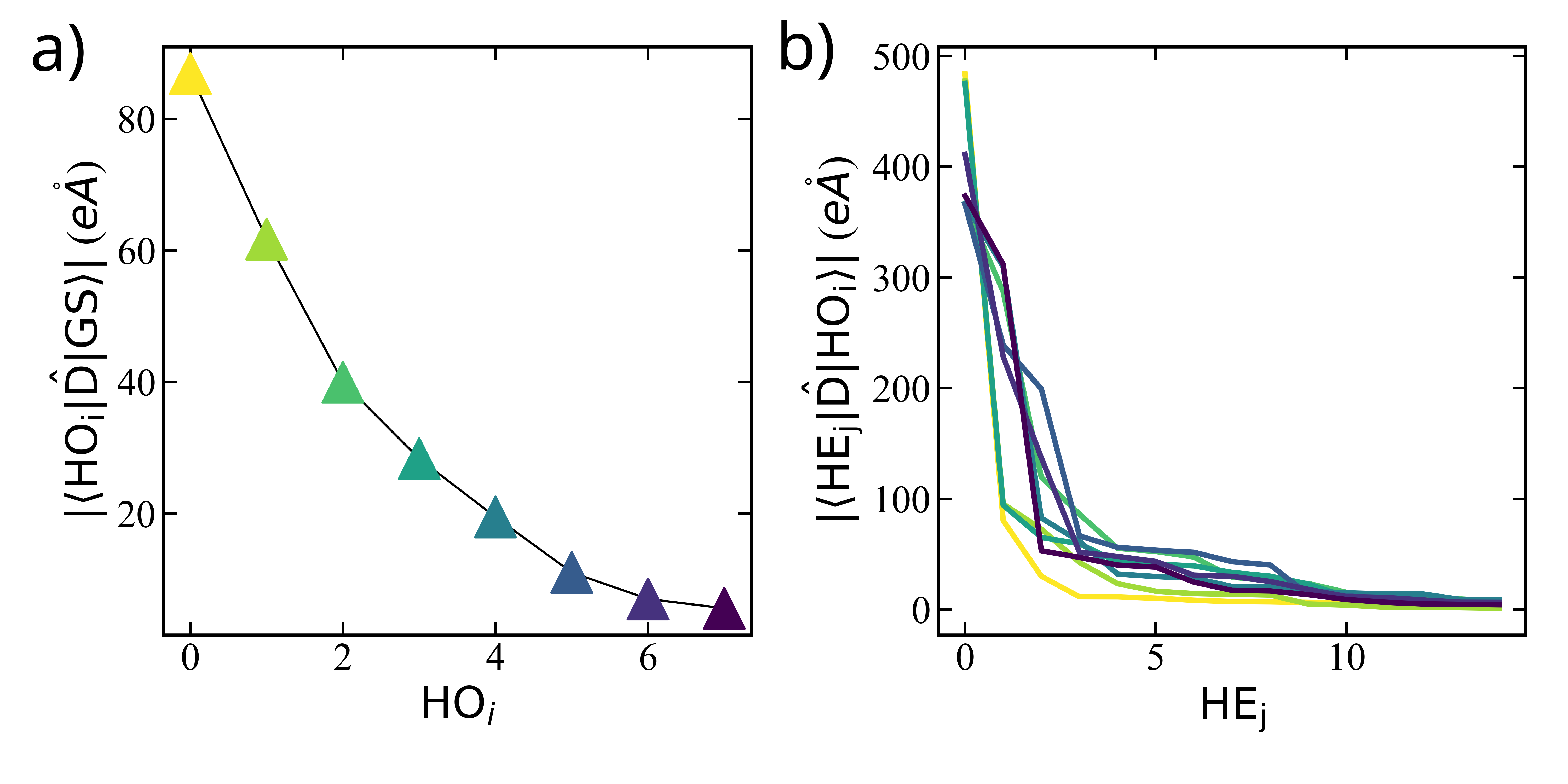}
    \caption{
    \textbf{Dipole couplings in the two-buckyball system.}
    \textbf{(a)} Dipole couplings between the ground state (GS) and high-energy odd-parity states (HO).
    \textbf{(b)} Dipole couplings between high-energy odd-parity states (HO) and high-energy even-parity states (HE), shown in descending order.
    }
    \label{fig:two-bucky-couplings}
\end{figure}

The rapid decay of the dipole couplings in Fig.~\ref{fig:two-bucky-couplings} provides a natural criterion for truncating the effective basis. We illustrate the effect of this truncation on the pairing dynamics of the buckyball dimer using three different bases: the full Hilbert space, a two-state model $\{|\mathrm{GS}\rangle,|\mathrm{HO}_0\rangle\}$, and a three-state model $\{|\mathrm{GS}\rangle,|\mathrm{HO}_0\rangle,|\mathrm{HE}_0\rangle\}$. Here, $|\mathrm{HO}_0\rangle$ is the odd-parity eigenstate with the largest dipole matrix element connecting it to the ground state, and $|\mathrm{HE}_0\rangle$ is the even-parity eigenstate with the largest dipole matrix element connecting it to $|\mathrm{HO}_0\rangle$.

\begin{figure}[h]
    \centering
    \includegraphics[width=0.9\textwidth]{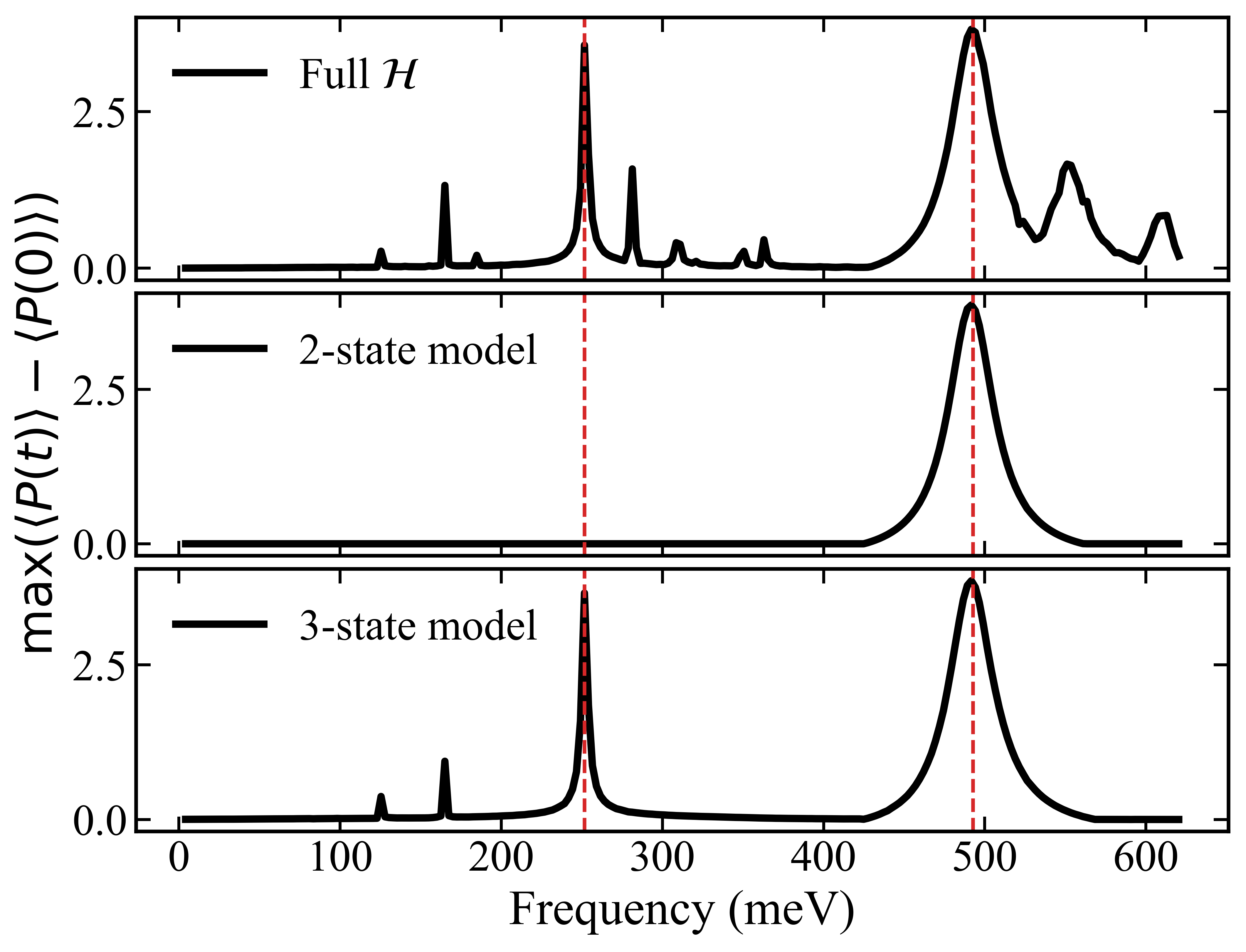}
    \caption{
    Maximum pair correlation for the driven two-buckyball system, comparing the full Hilbert space with truncated models. The three-state model successfully captures the resonant two-photon process at $h\nu\sim U/2$.
    }
    \label{fig:twobuckpair}
\end{figure}

The full time-evolution results for the different bases are shown in Fig.~\ref{fig:twobuckpair}. The two-state model captures the broad peak corresponding to a one-photon resonant transition, while the three-state model also captures the sub-$U$ two-photon process. Remarkably, this drastic truncation of the Hilbert space---from $\dim(\mathcal{H})=400$ to only three states---quantitatively reproduces the exact pairing dynamics. As detailed in Sec.~\ref{sec:DMRG+Krylov}, this physically motivated compression scheme can be scaled efficiently to construct reduced bases for larger clusters, allowing us to bypass the severe computational bottlenecks of standard time-evolution methods.

\section{Second-order transition amplitude for a three-state driven Hamiltonian}
\label{sec:FGR}

Consider a three-state model whose equilibrium Hamiltonian is
\begin{equation}
    H_0
    =
    \mathrm{diag}
    \left(
    E_{\mathrm{GS}},
    E_{\mathrm{HO}},
    E_{\mathrm{HE}}
    \right)
\end{equation}
in the basis
\begin{equation}
    B
    =
    \{
    |\mathrm{GS}\rangle,
    |\mathrm{HO}\rangle,
    |\mathrm{HE}\rangle
    \},
\end{equation}
with parity eigenvalues $\{+1,-1,+1\}$, respectively. We add a time-dependent coupling that connects only states of opposite parity:
\begin{equation}
V(t)
=
\begin{pmatrix}
0 & ge^{i\omega t} & 0\\
ge^{-i\omega t} & 0 & g'e^{i\omega t}\\
0 & g'e^{-i\omega t} & 0
\end{pmatrix}.
\end{equation}

We calculate the transition amplitude from $|\mathrm{GS}\rangle$ to $|\mathrm{HE}\rangle$ in time-dependent perturbation theory. In the interaction picture,
\begin{equation}
    |\widetilde{\Psi}(t)\rangle
    =
    e^{iH_0t}|\Psi(t)\rangle,
\end{equation}
and
\begin{equation}
    \widetilde{V}(t)
    =
    e^{iH_0t}V(t)e^{-iH_0t}.
\end{equation}
This gives
\begin{equation}
\widetilde{V}(t)
=
\begin{pmatrix}
0
&
g e^{-it(E_{\mathrm{HO}}-E_{\mathrm{GS}}-\omega)}
&
0
\\
g e^{it(E_{\mathrm{HO}}-E_{\mathrm{GS}}-\omega)}
&
0
&
g'e^{-it(E_{\mathrm{HE}}-E_{\mathrm{HO}}-\omega)}
\\
0
&
g'e^{it(E_{\mathrm{HE}}-E_{\mathrm{HO}}-\omega)}
&
0
\end{pmatrix}.
\end{equation}

The transition amplitude is
\begin{equation}
    C_{\mathrm{HE}}(t,\omega)
    =
    \langle\mathrm{HE}|
    U_I(t,0)
    |\mathrm{GS}\rangle.
\end{equation}
Expanding the interaction-picture time-evolution operator to second order,
\begin{equation}
    \begin{split}
        U_I(t,0)
        \simeq{}&
        \mathbb{1}
        -i\int_0^t dt_1\,\widetilde{V}(t_1)
        \\
        &-
        \int_0^t dt_1
        \int_0^{t_1}dt_2\,
        \widetilde{V}(t_1)\widetilde{V}(t_2).
    \end{split}
\end{equation}
The zeroth-order contribution vanishes because
$\langle\mathrm{HE}|\mathrm{GS}\rangle=0$, and the first-order contribution vanishes because
$\langle\mathrm{HE}|\widetilde{V}(t)|\mathrm{GS}\rangle=0$. The leading contribution is therefore second order:
\begin{equation}
    \begin{split}
        C_{\mathrm{HE}}(t,\omega)
        ={}&
        -
        \int_0^t dt_1
        \int_0^{t_1}dt_2
        \\
        &\times
        \langle\mathrm{HE}|
        \widetilde{V}(t_1)
        |\mathrm{HO}\rangle
        \langle\mathrm{HO}|
        \widetilde{V}(t_2)
        |\mathrm{GS}\rangle.
    \end{split}
\end{equation}

Defining
\begin{equation}
    \Delta_1
    =
    E_{\mathrm{HO}}-E_{\mathrm{GS}}-\omega
\end{equation}
and
\begin{equation}
    \Delta_2
    =
    E_{\mathrm{HE}}-E_{\mathrm{HO}}-\omega,
\end{equation}
we obtain
\begin{equation}
    C_{\mathrm{HE}}(t,\omega)
    =
    -gg'
    \int_0^t dt_1\,
    e^{i\Delta_2t_1}
    \int_0^{t_1}dt_2\,
    e^{i\Delta_1t_2}.
\end{equation}
Performing the integrals gives
\begin{equation}
    \begin{split}
        C_{\mathrm{HE}}(t,\omega)
        ={}&
        \frac{gg'}{\Delta_1}
        \left[
        \frac{
        e^{it(\Delta_1+\Delta_2)}-1
        }{
        \Delta_1+\Delta_2
        }
        -
        \frac{
        e^{it\Delta_2}-1
        }{
        \Delta_2
        }
        \right]
        \\
        ={}&
        \frac{gg'}{
        E_{\mathrm{HO}}-E_{\mathrm{GS}}-\omega
        }
        \\
        &\times
        \left[
        \frac{
        e^{it(E_{\mathrm{HE}}-E_{\mathrm{GS}}-2\omega)}-1
        }{
        E_{\mathrm{HE}}-E_{\mathrm{GS}}-2\omega
        }
        \right.
        \\
        &\left.
        \qquad
        -
        \frac{
        e^{it(E_{\mathrm{HE}}-E_{\mathrm{HO}}-\omega)}-1
        }{
        E_{\mathrm{HE}}-E_{\mathrm{HO}}-\omega
        }
        \right].
    \end{split}
    \label{eq:transition_amplitude}
\end{equation}

For detunings such that none of the denominators in Eq.~\ref{eq:transition_amplitude} vanish, $C_{\mathrm{HE}}(t,\omega)$ remains bounded as $t\to\infty$. The relevant resonant frequencies are
\begin{equation}
    \omega
    =
    E_{\mathrm{HO}}-E_{\mathrm{GS}},
\end{equation}
\begin{equation}
    \omega
    =
    E_{\mathrm{HE}}-E_{\mathrm{HO}},
\end{equation}
and
\begin{equation}
    2\omega
    =
    E_{\mathrm{HE}}-E_{\mathrm{GS}}.
\end{equation}
To evaluate the resonant limits, we use
\begin{equation}
    \lim_{x\to0}
    \frac{e^{itx}-1}{x}
    =
    it.
\end{equation}
Thus, when a resonance condition is satisfied, the corresponding contribution to the transition amplitude grows linearly in time. In particular,
\begin{equation}
    \omega
    =
    \frac{
    E_{\mathrm{HE}}-E_{\mathrm{GS}}
    }{2}
\end{equation}
is the two-photon resonance: two photons, each carrying half of the GS--HE energy gap, drive the transition into the high-energy even-parity state through the intermediate odd-parity state.

\section{DMRG+Krylov algorithm}
\label{sec:DMRG+Krylov}

We aim to compute the Hamiltonian dynamics of large K$_3$C$_{60}$ clusters, for which exact diagonalization is no longer computationally feasible. We adopt a matrix-product-state (MPS)-based approach because it provides a natural control parameter, the bond dimension $\chi$, for tuning the desired accuracy. Although several established MPS time-evolution schemes exist, including TEBD and TDVP, the three-orbital and three-dimensional nature of our system induces long-range entanglement in the MPS representation, making these methods prohibitively expensive. Instead, we exploit the coupling structure of the driving term and the symmetry constraints of the lattice, guided by our exact-diagonalization results for the buckyball dimer; see Sec.~\ref{sec:two-bucky-ED}.

The algorithm constructs an effective Hamiltonian retaining the ground state, a set of $n_{\mathrm{HO}}$ high-energy odd-parity states that couple most strongly to the ground state, and, for each HO state, a corresponding high-energy even-parity block containing the $n_{\mathrm{HE}}$ even states with the largest couplings to that HO state.

First, we choose the buckyball-cluster geometry and shift the coordinate origin so that inversion maps the cluster onto itself and parity is well defined. Without loss of generality, we take the drive direction to be $x$. We perform a DMRG calculation to obtain an MPS representation of the ground state $|\mathrm{GS}\rangle$ with a chosen maximum bond dimension $\chi$, and construct the dipole operator $D^x$ as a matrix-product operator.

To build the HO subspace, we run a Lanczos procedure seeded with
\begin{equation}
    |\mathrm{DGS}\rangle
    =
    D^x|\mathrm{GS}\rangle,
\end{equation}
generating the odd-parity Krylov subspace
\begin{equation}
    \begin{split}
        \mathcal{K}^{\mathrm{odd}}_n
        \left(
        H_0,
        |\mathrm{DGS}\rangle
        \right)
        =
        \{
        &|\mathrm{DGS}\rangle,
        H_0|\mathrm{DGS}\rangle,
        \ldots,
        \\
        &H_0^n|\mathrm{DGS}\rangle
        \}.
    \end{split}
\end{equation}
This space contains only odd-parity states because the Hamiltonian preserves parity. Diagonalizing $H_0$ within $\mathcal{K}^{\mathrm{odd}}_n$ yields $n+1$ approximate eigenstates. From these, we select the $n_{\mathrm{HO}}$ states with the largest dipole couplings to the ground state.

For each selected state $|\mathrm{HO}_i\rangle$, we seed a second Lanczos run with
\begin{equation}
    |\mathrm{DHO}_i\rangle
    =
    D^x|\mathrm{HO}_i\rangle
\end{equation}
to generate the even-parity Krylov subspace
\begin{equation}
    \begin{split}
        \mathcal{K}^{\mathrm{even}}_n
        \left(
        H_0,
        |\mathrm{DHO}_i\rangle
        \right)
        =
        \{
        &|\mathrm{DHO}_i\rangle,
        H_0|\mathrm{DHO}_i\rangle,
        \ldots,
        \\
        &H_0^n|\mathrm{DHO}_i\rangle
        \}.
    \end{split}
\end{equation}
Every vector in this space has even parity. The Lanczos construction preferentially resolves eigenstates having large overlap with $|\mathrm{DHO}_i\rangle$. Diagonalizing the Hamiltonian in this even subspace also recovers the ground state, which is removed from the final excited-state basis.

Even states generated from different HO seeds are generally not mutually orthogonal. We therefore track their overlap matrix and orthogonalize the union of the even-sector basis states before constructing the final reduced representation. This prevents duplicate or nearly duplicate HE states from being counted multiple times.

We then construct the equilibrium Hamiltonian, dipole operator, and pair-correlation operator in the effective basis
\begin{equation}
    \begin{split}
        B_{\mathrm{eff}}
        =
        \{
        |\mathrm{GS}\rangle
        \}
        \cup
        \{
        |\mathrm{HO}_i\rangle
        \}
        \cup
        \{
        |\mathrm{HE}^{(i)}_j\rangle
        \},
    \end{split}
\end{equation}
where $i=1,\ldots,n_{\mathrm{HO}}$ and $j=1,\ldots,n_{\mathrm{HE}}$. Before removal of linearly dependent vectors, the nominal basis dimension is
\begin{equation}
    \dim(B_{\mathrm{eff}})
    =
    1+n_{\mathrm{HO}}
    +n_{\mathrm{HO}}n_{\mathrm{HE}}.
\end{equation}

In practice, we use 128 Lanczos steps per run. Depending on the geometry, we retain between three and ten HO states and 15 HE states per HO state. To mitigate loss of orthogonality, we implement the coefficient-based reorthogonalization method described in Ref.~\cite{dargel_2012}, which minimizes the number of MPS additions required. Finally, we propagate the reduced system to obtain the dynamics and pair expectation values. The computational bottleneck lies entirely in constructing the effective basis; the subsequent time evolution has negligible computational cost. The construction targets the lowest nonvanishing optical orders and is therefore best controlled for weak electric fields. Its accuracy can be improved systematically by including more odd- and even-parity states and, if required, additional alternating-parity Krylov layers.

\section{Ground-state parity for the one-orbital 14-site 3D cluster}
\label{sec:sup_parity}

For the 14-site one-orbital cluster, the ground-state parity changes from even to odd at $U/W\gtrsim1.16$. This parity inversion is unique to this specific cluster geometry among those studied. A gap opens between the odd ground state and the lowest even state, reaching a maximum of approximately \SI{4}{\milli\electronvolt}. The calculations presented in the main text are initialized from the lowest even-parity state.

\begin{figure}[h]
    \centering
    \includegraphics[width=0.9\textwidth]{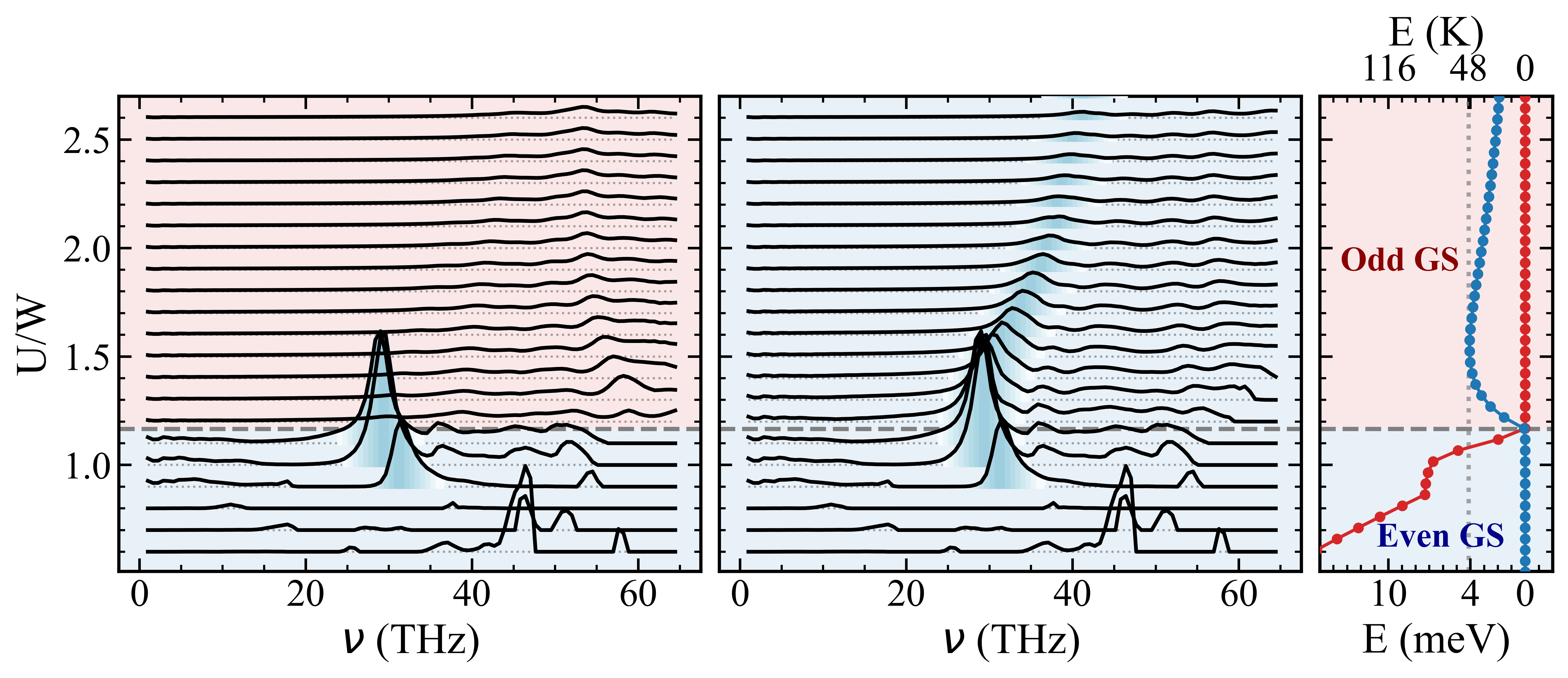}
    \caption{
    Light-induced pair correlations for the 14-site one-orbital cluster, comparing calculations initialized from the true ground state (left) and the lowest-energy even-parity state (middle). A ground-state parity transition occurs at $U/W\approx1.16$ (right).
    }
    \label{fig:true_gs}
\end{figure}

Under experimental conditions, both the odd- and even-parity states can carry nonvanishing thermal weight because the experimental temperatures exceed this small energy splitting. Figure~\ref{fig:true_gs} compares the light-induced pair correlations obtained when initializing the system in the true ground state with those obtained from the lowest-energy even state. Initializing the dynamics in the odd state suppresses the resonant enhancement of pair correlations. This suppression occurs because the particular high-energy state responsible for the enhanced pair correlations has even parity and is therefore inaccessible through a two-photon process from an odd initial state under the inversion selection rules. Consequently, a thermal mixture can still exhibit light-induced pair correlations, but their overall intensity is reduced by the population residing in the odd sector. For the $U/W$ regime most relevant to K$_3$C$_{60}$, however, the ground state retains even parity in all clusters investigated.

\section{Two-photon absorption}
\label{sec:two_phto_ab_sup}

Here, we calculate the two-photon absorption as a function of $U/W$ by evaluating the correlator
\begin{equation}
    \begin{split}
        G_{D^2}
        =
        \langle\mathrm{GS}|
        \hat{D}^2
        \frac{1}{
        2h\nu+E_0-\hat{H}_0+i\eta
        }
        \hat{D}^2
        |\mathrm{GS}\rangle.
    \end{split}
\end{equation}
This expression provides an approximation to Eq.~\ref{eq:two-phot} of the main text in the frequency regime where the intermediate one-photon transitions are sufficiently off-resonant that their energy dependence can be absorbed into an effective $D^2$ vertex.

\begin{figure}[h]
    \centering
    \includegraphics[width=0.9\textwidth]{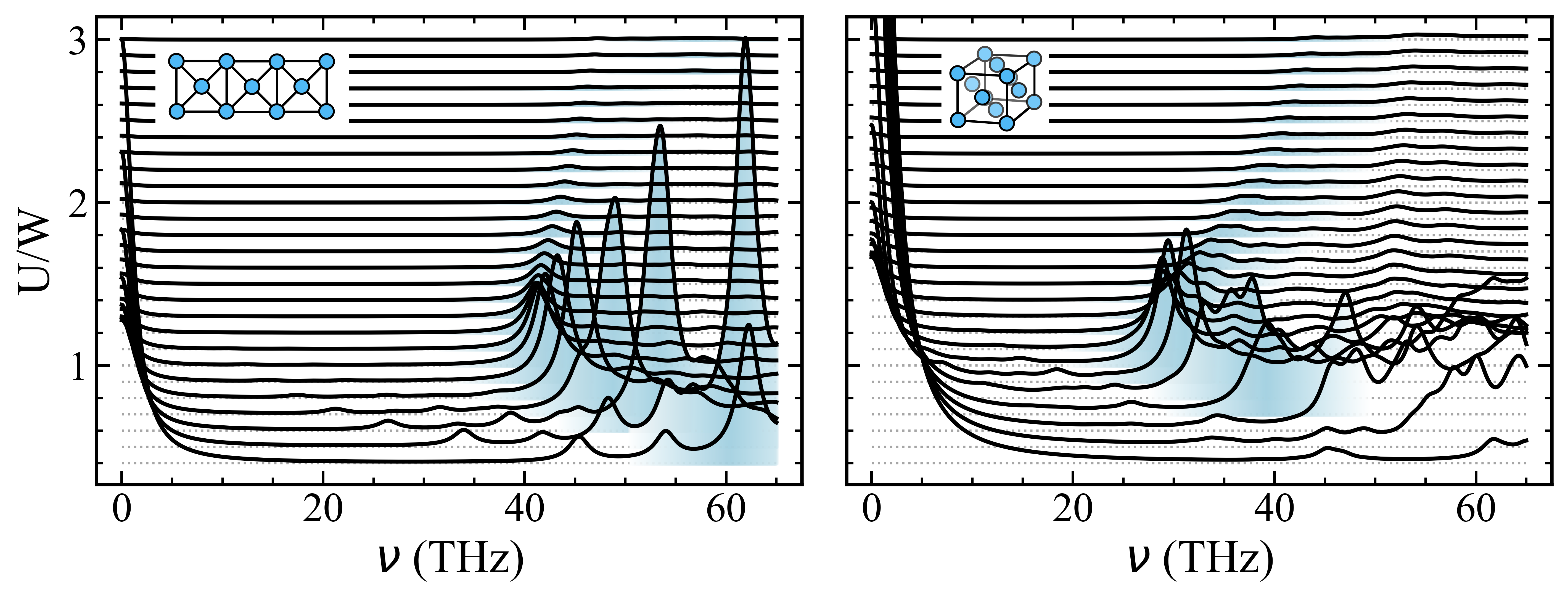}
    \caption{
    Two-photon absorption for the 11-site 2D cluster (left) and the 14-site 3D one-orbital cluster (right). The highlighted peak corresponds to the GS$\to$HO$\to$HE two-photon pathway described in the main text.
    }
    \label{fig:two_ph_absorption}
\end{figure}

The calculated two-photon absorption for the one-orbital 11-site 2D and 14-site 3D clusters is presented in Fig.~\ref{fig:two_ph_absorption}. The finite amplitude at zero frequency contains an elastic contribution associated with virtual excitation and de-excitation back into the ground state. The finite-frequency features of the two-photon absorption closely follow the resonant enhancement of pair correlations discussed in the main text.

It is important, however, to distinguish the two-photon absorption strength from the pair correlations of the final state. Whereas the two-photon absorption amplitude decreases approximately monotonically with increasing $U/W$, the light-induced pair-correlation enhancement reaches a maximum near $U/W\approx1$. This indicates that the intermediate-coupling regime achieves an optimal balance between optical access to the resonant HE state and the intrinsic pairing character of that state.

\end{document}